\documentclass[apjfonts]{emulateapj}
\usepackage{appendix,natbib}
\usepackage{amsmath}
\usepackage{courier}

\usepackage[pdfpagemode=UseNone,pdfstartview=FitH,colorlinks=true,citecolor=blue,linkcolor=blue,urlcolor=blue]{hyperref}
\usepackage[all]{hypcap}

\newcommand{\halpha}{H\ensuremath{\alpha}}
\newcommand{\hbeta}{H\ensuremath{\beta}}
\newcommand{\um}{\ensuremath{\mu}m}
\newcommand{\kmps}{km s\ensuremath{^{-1}}}
\def\msun{{\rm\,M_\odot}}

\begin{document}

\title{The MOSDEF survey: the prevalence and properties of galaxy-wide AGN-driven outflows at $\lowercase{z}\sim 2$}

\author{\sc Gene C. K. Leung\altaffilmark{1}, 
Alison L. Coil\altaffilmark{1}, 
Mojegan Azadi\altaffilmark{1},
James Aird\altaffilmark{2},
Alice Shapley\altaffilmark{3}, 
Mariska Kriek\altaffilmark{4}, 
Bahram Mobasher\altaffilmark{5}, 
Naveen Reddy\altaffilmark{5}, 
Brian Siana\altaffilmark{5},
William R. Freeman\altaffilmark{5}, 
Sedona H. Price\altaffilmark{4}, 
Ryan L. Sanders\altaffilmark{3}, 
Irene Shivaei\altaffilmark{5}}

\altaffiltext{1}{Center for Astrophysics and Space Sciences, University of California, San Diego, La Jolla, CA 92093, USA}
\altaffiltext{2}{Institute of Astronomy, University of Cambridge, Madingley Road, Cambridge CB3 0HA, UK}
\altaffiltext{4}{Astronomy Department, University of California, Berkeley, CA 94720, USA}
\altaffiltext{3}{Department of Physics \& Astronomy, University of California, Los Angeles, CA 90095, USA}
\altaffiltext{5}{Department of Physics \& Astronomy, University of California, Riverside, CA 92521, USA}

\begin{abstract}

Using observations from the first two years of the MOSFIRE Deep Evolution Field (MOSDEF) survey, we study 13 AGN-driven outflows detected from a sample of 67 X-ray, IR and/or optically-selected AGN at $z \sim 2$.  
The AGN have bolometric luminosities of $\sim10^{44}-10^{46} ~\mathrm{erg~s^{-1}}$, including both quasars and moderate-luminosity AGN.
We detect blueshifted, ionized gas outflows in the \hbeta , [OIII], \halpha ~and/or [NII] emission lines of 19\% of the AGN, while only 1.8\% of the MOSDEF galaxies have similarly-detected outflows. 
The outflow velocities span $\sim$300 to 1000 km s$^{-1}$.  Eight of the 13 outflows are spatially extended on similar scales as the host galaxies, with spatial extents of 2.5 to 11.0 kpc. 
Outflows are detected uniformly across the star-forming main sequence, showing little trend with the host galaxy SFR. 
Line ratio diagnostics indicate that the outflowing gas is photoionized by the AGN. 
We do not find evidence for positive AGN feedback, in either our small MOSDEF sample or a much larger SDSS sample, using the BPT diagram. 
Given that a galaxy with an AGN is ten times more likely to have a detected outflow, the outflowing gas is photoionzed by the AGN, and estimates of the mass and energy outflow rates indicate that stellar feedback is insufficient to drive at least some of these outflows, they are very likely to be AGN-driven. 
The outflows have mass-loading factors of the order of unity, suggesting that they help regulate star formation in their host galaxies, though they may be insufficient to fully quench it.

\end{abstract}

\maketitle

\section{Introduction}
Perhaps one of the most remarkable discoveries in astronomy over the last two decades is that supermassive black holes (mass greater than $\sim10^6~M_\odot$) exist in the nuclei of virtually all galaxies \citep[e.g.][]{mag98, hec14}. 
Supermassive black holes grow by accreting gas from the centers of their galaxies, which are observed as active galactic nuclei \citep[AGN, e.g.,][]{ant93, net15}.
The growth rates of supermassive black holes and galaxies over cosmic time, traced by black hole accretion density and star formation density respectively, follow a strikingly similar pattern for the last $\sim11$ billion years \citep[e.g.][]{ued03, hop06b, aird15}.
Furthermore, a tight relation exists between the masses of central supermassive black holes and various properties of their galactic bulges \citep[e.g.,][]{fer00, geb00}.
Such empirical evidence clearly indicates a co-evolution between galaxies and their supermassive black holes.

An important theoretical connection between AGN and galaxies is inferred from the comparison between models of galaxy formation and observed galaxy properties.
The observed stellar mass function of galaxies, which traces the distribution of galaxies by stellar mass, shows a sharp cutoff at the high mass end relative to the dark matter halo mass function.
Without some kind of feedback mechanism(s) releasing energy into the galaxy in the theoretical models, strong gas cooling in massive, luminous galaxies would lead to more galaxies being formed at the high mass / high luminosity end of the luminosity function than are observed \citep[e.g.,][]{ben03}.

Most of the current successful models of galaxy formation \citep[e.g.][]{dm05, hop06a, hop08, deb12} invoke AGN as the primary source of feedback in massive galaxies, in order to inject energy into the interstellar medium, regulate or quench star formation, and slow black hole growth. 
A key process for AGN feedback is AGN-driven outflows, in which the AGN produces a high-velocity wind that heats or sweeps up gas and drives them into the interstellar medium over distances comparable to the size of the galaxy, leading to the regulation and/or quenching of star formation.
Although these models successfully produce the observed galaxy stellar mass function, they must be verified with observations.
Therefore, it is a critical area of on-going research to observe and characterize AGN-driven outflows.

It is now evident that AGN-driven outflows are common in the local Universe ($z\lesssim 1$). 
Ionized outflows have long been known to manifest themselves in a blueshifted wing in the emission lines of AGN spectra \citep[e.g.][]{wee70}.
The prevalence of these outflows are revealed by large spectroscopic surveys such as the Sloan Digital Sky Survey (SDSS).
Using $>24000$ AGN ($L_\mathrm{[OIII]} \sim 10^{40} - 10^{44}\mathrm{~erg~s}^{-1}$) from SDSS, \citet{mul13} find signs of ionized outflows in the [OIII]$\lambda5007$ emission line profile in $17 \%$ of AGN, with outflow velocities between 600 and 2000 \kmps .
\citet{zak14} also find widespread ionized outflows in $>$500 luminous AGN ($L_\mathrm{[OIII]} \sim 10^{42} - 10^{43}\mathrm{~erg~s}^{-1}$) in SDSS, with outflow velocities of 400 to 4800 \kmps . 

The fiber spectroscopy provided by SDSS does not allow for measurements of the physical extent of the 
outflows, which can be determined by 
follow-up integral field spectroscopy (IFS) observations. 
Using IFS, the physical extent of the outflows in very luminous AGN ($L_\mathrm{[OIII]} \gtrsim 10^{42}\mathrm{~erg~s}^{-1}$) are shown to be $\sim$ 6 to 19 kpc, comparable to the size of the host galaxies \citep[e.g.][]{liu13, har14}.
These studies show that galaxy-wide ionized outflows are relatively common among the most luminous AGN in the local Universe.

However, the crucial epoch to observe AGN-driven outflows is $z\sim 2$, when the Universe was only a quarter of its current age. 
The cosmic black hole accretion and star formation densities both reach a maximum at $z\sim 2$, before gradually declining by about a factor of 10 to the present-day values \citep[e.g.][]{aird15}.
Not only are AGN-driven outflows expected to be more prevalent at this epoch of peak black hole accretion activity, their relation to the subsequent decline of cosmic star formation density is of particular interest in understanding the role of AGN feedback in the regulation and quenching of star formation.
Therefore, it is a key cosmic time to observe and characterize AGN-driven outflows and investigate their impact on their host galaxies.

Observations of AGN-driven outflows at this epoch of peak activity level in the Universe remain limited. 
Many of the studies in the existing literature at $z\sim 2$ are focused on small samples of extreme
sources.
Some early results include \citet{nes08}, which finds outflows in three powerful radio-loud AGN in very massive galaxies.
\citet{har12} finds energetic galaxy-wide outflows in a sample of 8 $z\sim 2$ AGN-hosting ultraluminous infrared galaxies (ULIRGs), which are galaxies with elevated star formation activity.
Powerful outflows have also been found in a study of 10 $z \sim 1.5$ luminous quasars \citep{bru15, per15}, which are the most luminous members of the AGN population ($L_\mathrm{bol}\gtrsim 10^{45} \mathrm{~erg~s}^{-1}$).
These studies consist of special subclasses of AGN, which are not representative of the much larger AGN population.
In order to investigate how AGN feedback affects galaxy evolution in the wider picture, a study using a more complete sample of AGN is necessary.

Recent studies using larger samples of a few tens of AGN report evidence of prevalent AGN-driven outflows.
\citet{har16} study [OIII] in a sample of 54 X-ray selected AGN ($L_\mathrm{X} \sim  10^{42} -  10^{45}\mathrm{~erg~s}^{-1}$) at somewhat lower redshifts of $z \sim 1.1-1.7$ and find that a 
second kinematic component is detected in $35 \%$ of the [OIII]-detected AGN, while $\sim 50\%$ have line widths $>600$\kmps , with a maximum line width of $\sim$1000 \kmps .
The largest sample at $z \sim 2$ to date is from \citet{gen14}, where the \halpha ~and [NII] line profiles 
are used to find signs of outflows in the nuclear region (radius $<$ 2.5 kpc) of 34 out of 110 star-forming galaxies at $z\sim 1-3$, 18 of which are confirmed AGN ($L_\mathrm{AGN} \sim  10^{44} -  10^{46}\mathrm{~erg~s}^{-1}$).
With outflow velocities of 300 to 1300 \kmps , the estimated mass outflow rates (though very uncertain) are comparable to the star formation rates (SFR) of the host galaxies, suggesting the outflows may be capable of ejecting sufficient gas to quench star formation in the host galaxies. 

It is important to test these results with a larger and more representative sample of AGN at $z\sim2$.
Moreover, to better constrain and characterize the kinematics of the outflows, it is important to have simultaneous measurement of \hbeta , [OIII], \halpha ~and [NII]; [OIII] is especially useful, as it is unlikely affected by any potential AGN broad line emission.  The commissioning of the MOSFIRE multi-object spectrograph \citep{mcl10, mcl12} on the 10-meter Keck I telescope provides great opportunities to systematically study AGN-driven outflows at high redshift.
It is now possible to simultaneously obtain rest-frame optical spectra for a large number of sources at high redshift.
In this study, we make use of data from the MOSFIRE Deep Evolution Field (MOSDEF) survey \citep{kri15}, a recently-completed four year near-infrared spectroscopic survey using MOSFIRE.
With MOSDEF, we have obtained rest-frame optical spectra of $\sim 1500$ galaxies in $1.4 < z < 3.8$, including more than 100 AGN.
These rest-frame optical spectra cover important emission lines for characterising AGN and AGN-driven outflows, including \hbeta , [OIII], \halpha ~and [NII].
The spectroscopic targets in the survey are selected to a fixed H-band magnitude, roughly corresponding to a threshold in stellar mass.
These selection criteria result in a more representative sample of galaxies that is not biased towards high SFRs.
Moreover, AGN in this study are identified using multi-wavelength data (see Section \ref{data}), which minimizes AGN selection biases \citep{coil15,aza16}.
This provides us with a large sample consisting of more typical AGN in more representative galaxies, allowing us to investigate the effects of AGN-driven outflows in the general AGN population.

In this paper we present the results using data from the first two years of the MOSDEF survey.
In Section \ref{data} we describe the survey and our AGN sample; in Section \ref{results} we report our detection and analysis of outflows among our AGN; in Section \ref{discuss} we discuss the implications of our results. We conclude in Section \ref{conc}.

\section{Observations and AGN Sample}\label{data}
In this section we describe the dataset used in this study and the methods employed to identify AGN at various wavelengths.
We also outline how we estimate host galaxy properties using SED fitting.

\subsection{The MOSDEF Survey}
In this study we use data from the first two years of the MOSDEF survey \citep{kri15}, which makes use of the MOSFIRE spectrograph \citep{mcl10, mcl12} on the Keck I telescope.
In the first two years of observations, a total of 713 targets were observed, 555 of which yielded a reliable redshift.
With the complete MOSDEF survey, we have obtained moderate-resolution ($R = 3000-3650$) rest-frame optical spectra of $\sim1500$ galaxies and AGN at $1.4 \lesssim z \lesssim 3.8$ in three well-studied CANDELS fields \citep{gro11, koe11}: AEGIS, COSMOS and GOODS-N. 
Data were obtained for three masks in two additional CANDELS fields, GOODS-S and UDS, because our primary target fields were not visible during a portion of a few of the observed nights.
All of the MOSDEF targets are covered by the 3D-{\it HST} survey \citep{skel14}.
Full technical details of the design and observations of the MOSDEF survey can be found in \citet{kri15}.
MOSDEF targets are selected within three redshift intervals, $1.37 \leq z \leq 1.70$, $2.09 \leq z \leq 2.61$ and $2.95 \leq z \leq 3.80$, such that the brightest rest-frame optical emission lines lie within atmospheric windows.
This allows coverage of the \hbeta, [OIII], \halpha ~and [NII] emission lines in the lower and middle redshift intervals, and \hbeta ~and [OIII] in the higher redshift interval.

Targets are prioritized according to weights determined by their H-band (F160W) magnitudes and redshifts in the 3D-{\it HST} catalogs \citep{skel14}, with emphasis given to brighter galaxies and galaxies within the middle redshift interval ($2.09 \leq z \leq 2.61$).
Targets are selected down to limiting H-band magnitudes of 24.0, 24.5 and 25.0 for the lower, middle and higher redshift intervals, respectively.
We assign higher targeting priority to galaxies that are known to host AGN, identified in advance by X-ray emission and/or red IRAC colors.
In addition, AGN are identified in the MOSDEF spectra using rest-frame optical diagnostics.
The identification of AGN in the MOSDEF survey is described below; full details are discussed in \citet{aza16}.

\subsection{X-ray AGN Sample}\label{x-ray}
X-ray AGN were identified prior to the MOSDEF slitmasks design using deep {\it Chandra} X-ray imaging in 
the survey fields.
The depth of the {\it Chandra} data used is 160 ks in COSMOS, 2 Ms in GOODS-N, 4 Ms in GOODS-S and 800 ks in AEGIS.
These correspond to hard band (2-10 keV) flux limits of $1.8\times 10^{-15}$, $2.8\times 10^{-16}$, $1.6\times 10^{-16}$ and $5.0\times 10^{-16}$ ergs s$^{-1}$ cm$^{-2}$, respectively.

The X-ray data from all the fields were reduced and analyzed according to the prescription in \citet[][see also \citealt{geo14, nan15}]{lai09}.
For point source detection, a false probability threshold of $4\times 10^{-6}$ was adopted. 
Each source must be detected in at least one of the four bands: soft (0.5-2 keV), hard (2-7 keV), ultrahard (4-7 keV) and full (0.5-7 keV).
Then the X-ray sources were matched to sources detected at IRAC, NIR and optical wavelengths using the likelihood ratio method \citep[see][]{cil03, cil05, bru07, luo10} to identify secure multiwavelength counterparts. 
(See \citealt{aird15} for details.)
The resulting catalogs were then matched to the closest source within $1''$ in the 3D-{\it HST} catalogs used for MOSDEF target selection.
These X-ray sources were given higher priority in MOSDEF target selection.

For the X-ray sources observed by MOSDEF and for which a redshift was determined, 
the 2-10 keV rest-frame X-ray luminosity was estimated from the hard or soft band fluxes.
Luminosity estimates assume a simple power-law spectrum with photon index $\Gamma = 1.9$ and Galactic absorption only.
Our hard band detection threshold approximately corresponds to an X-ray luminosity limit of $L_\mathrm{2-10 keV} \approx 1.3-15.1 \times 10^{42} \mathrm{~erg~s}^{-1}$ across the four fields.
The luminosity estimates should be reasonable since at the MOSDEF redshift range ($z > 1.4$), a relatively large absorption column density ($N_H \gtrsim 10^{23} \mathrm{~cm}^{-2}$) is required to significantly suppress the observed flux, even in the soft band.
Of the 28 X-ray AGN observed in the first two years of the MOSDEF survey with a secure derived redshift, 22 are detected in the hard band, with $L_\mathrm{2-10 keV}$ spanning $\sim 10^{43}-10^{45} \mathrm{~erg~s}^{-1}$.
All 28 X-ray AGN are included in this study.

\subsection{IR AGN Sample}\label{ir}
Although deep X-ray surveys are a highly reliable means of selecting AGN, X-ray photons can be significantly absorbed at high column densities (e.g. $N_H \gtrsim 10^{23} \mathrm{~cm}^{-2}$), weakening their efficiency in identifying heavily obscured AGN. 
The MIR wavelengths can potentially be used to identify such AGN, as high-energy photons from the AGN are absorbed and re-radiated by dust at the MIR wavelengths.

It is common to use the unique MIR colors of AGN to identify infrared AGN (IR AGN) using IRAC data from {\it Spitzer} \citep[e.g.][]{lac04, ster05}.
For MOSDEF, we make use of the IRAC color criteria from \citet{don12} to select IR AGN.
These criteria were designed to limit contamination from star forming galaxies at high redshift ($z\sim3$), resulting in a pure AGN sample, and were used by \citet{aza16} for the MOSDEF survey.

We use IRAC fluxes reported in the 3D-{\it HST} catalogs \citep{skel14}, and select IR AGN following the \citet{don12} criteria in IRAC color-color space:
\begin{gather} 
x={\rm log_{10}}\left( \frac{f_{\rm 5.8 \um}}{f_{\rm 3.6 \um}}\right), 
\quad y={\rm log_{10}}\left( \frac{f_{\rm 8.0 \um}}{f_{\rm 4.5 \um}}\right) \\
x \ge 0.08 \textrm{~ and ~} y \ge
0.15\\ 
y \ge (1.21\times{x})-0.27 \label{eq:relax}\\ 
y \le (1.21\times{x})+0.27\\ 
f_{\rm 4.5\um} > f_{\rm 3.6 \um} \label{eq:pl1}\\
f_{\rm 5.8 \um} > f_{\rm 4.5 \um} \label{eq:pl2}\\ 
f_{\rm 8.0 \um} > f_{\rm 5.8 \um}\label{eq:pl3}. 
\end{gather}
Conditions \ref{eq:pl1}, \ref{eq:pl2} and \ref{eq:pl3} are slightly relaxed to include sources satisfying these conditions within 1$\sigma$ uncertainties of their IRAC fluxes.
Condition \ref{eq:relax} is also relaxed to include one source falling marginally outside this condition but satisfying all the others and having colors far from any star-forming galaxies in the MOSDEF sample.
27 IR AGN are selected in the data from the first two years of MOSDEF in this way; 9 of these IR AGN are also identified through X-ray imaging as described in Section \ref{x-ray}.

\subsection{Optical AGN Sample}\label{opt}

Apart from the 46 X-ray and/or IR AGN identified prior to target selection as described in Sections \ref{x-ray} and \ref{ir}, we additionally identify AGN optically using emission line ratios derived from the MOSDEF rest-frame optical spectra.
For the 555 MOSDEF sources from the first two years that have a reliable redshift, as described in \citet{aza16}, we simultaneously fit the \hbeta, [OIII], [NII] and \halpha ~emission lines using a $\chi^2$-minimization method by the \texttt{MPFIT} \citep{mar09} routine in \texttt{IDL}. 
For sources in the highest redshift interval of $2.95 \leq z \leq 3.80$, we fit only the \hbeta ~and [OIII] emission lines, as the spectra
do not cover the [NII] and \halpha ~lines.
We fit the spectra with a continuum near the emission lines with zero slope and a maximum of three Gaussian components for each line:
\begin{enumerate}
\item A Gaussian function with FWHM $<$ 2000 \kmps, one each for the \hbeta, [OIII], [NII] and \halpha ~emission lines, representing the narrow-line emission from the AGN and/or host galaxy;
\item A broad Gaussian function with FWHM $>$ 2000 \kmps, one each for \hbeta ~and \halpha ~only, representing the broad line emission from the AGN;
\item An additional Gaussian function with FWHM $<$ 2000 \kmps ~and a velocity offset between $-500$ and 0 \kmps , one each for the \hbeta, [OIII], [NII] and \halpha ~emission lines, representing a potential blueshifted outflow component.
\end{enumerate}
The FWHM of each Gaussian function is set  to be equal for all of the lines of interest.
The flux ratios between [OIII]$\lambda 4960$ and [OIII]$\lambda 5008$ and between [NII]$\lambda 6550$ and [NII]$\lambda 6585$ are fixed at 1:3 for each Gaussian function.
We determine the minimum width of the Gaussian functions by randomly selecting unblended sky lines in different wavelength intervals and fitting them with a single Gaussian function.
From the average width of the sky lines, we determine a minimum width (Gaussian sigma) in observed wavelength of 2.5\AA ~for \hbeta ~ and [OIII] and 3.5\AA ~for \halpha ~and [NII]. 
We first perform fits and evaluate the $\chi^2$ with a model consisting of the narrow-line component (component 1) alone, then only accept additional components when the resulting decrease in $\chi^2$ implies a $p$-value of $<0.01$ for the simpler model. 

Based on the results from the emission line fits, we use the diagnostic BPT diagram \citep{bal81, vei87} to identify optical AGN in the MOSDEF sample.  For this diagnostic we sum the non-broad component fluxes of the \hbeta, [OIII], [NII] and \halpha ~emission lines, applying Balmer absorption corrections to the \hbeta ~and \halpha ~fluxes.

Sources above the \citet{kau03} demarcation line in the BPT diagram are selected as optical AGN for the purposes of 
this study, as at low redshift sources above this line are considered likely to have an AGN contribution to their optical spectra.  
At high redshifts, star-forming galaxies are known to be offset from SDSS star-forming galaxies in the BPT diagram \citep[e.g.][]{shap15, san16}.
In the MOSDEF survey, this offset is observed only for galaxies in the lower stellar mass half of the sample ($M_* < 10^{10} \msun$) \citep{shap15}, while most of the AGN identified in the MOSDEF have stellar masses of $M_* > 10^{10} \msun$.
Previous MOSDEF AGN papers \citep{coil15, aza16} used the \citet{mel14} demarcation line in the BPT digram to identify optical AGN, as it balances contamination by star-forming galaxies and completeness of the AGN sample in our survey.
In this study, the aim is to select sources with at least some AGN contribution instead of pure AGN, as the presence of any AGN activity may suffice to drive an outflow and provide feedback.
Therefore, an AGN sample as complete as possible is preferred, even if it has some contamination.
The use of the \citet{kau03} line instead of the \citet{mel14} line results in nine additional sources being identified as potential AGN.
This relative increase in the total number of AGN is consistent with the initial MOSDEF sample presented in \citet{coil15},
who show that at $z\sim2$ sources in the BPT diagram on average are offset high by $\sim0.1$ dex compared to low redshift samples.
This small shift implies that the \citet{kau03} demarcation line should be fairly reliable for our sample, if possibly somewhat contaminated.

Using this criterion, we have selected 21 optical-only AGN in addition to the 46 AGN selected through X-ray and/or IR emission.  This results in a total of 67 X-ray, IR and/or optical AGN in our sample.  
Details of the overlap and uniqueness of AGN
identified at each wavelength in our sample are given in \citet{aza16}. 
The detection rates of outflows in AGN subsamples detected at different wavelengths are discussed below in Section \ref{ID}.  We note that only two of the 13 AGN for 
which we detect outflows (presented below) are identified as AGN solely using the BPT diagram.

\capstartfalse
\begin{deluxetable*}{llrrccccc}
\centering
\tablecaption{Outflow Host Properties}
\tablehead{
\colhead{ID}&\colhead{Field}&\colhead{R.A.}&
\colhead{Decl.}&\colhead{$z$}&
\colhead{Identification}&\colhead{SFR}&
\colhead{log($M_*/M_\odot$)}&
 \colhead{log($L_\mathrm{[OIII]}/\mathrm{erg~s}^{-1})$}
\\ \colhead{} & \colhead{} & \colhead{(J2000)} & \colhead{(J2000)}
 & \colhead{} & \colhead{Method} & \colhead{($M_\odot$ yr$^{-1}$)} & \colhead{}
& \colhead{}}
\startdata
6055	&	UDS	&	02:17:34	&	-05:14:48	&	2.3	&	IR	&	3	&	10.58	&	43.03	\\
20891	&	COSMOS	&	10:00:14	&	+02:22:57	&	2.19	&	Opt	&	129	&	10.55	&	43.38	\\
15286	&	COSMOS	&	10:00:15	&	+02:19:44	&	2.45	&	X-ray/IR/Opt	&	2399	&	10.63	&	42.39	\\
20979	&	COSMOS	&	10:00:16	&	+02:23:00	&	2.19	&	Opt	&	85	&	10.6	&	43.20	\\
11223	&	COSMOS	&	10:00:20	&	+02:17:25	&	2.1	&	X-ray/Opt	&	209	&	11.2	&	42.55	\\
9367	&	GOODS-N	&	12:37:04	&	+62:14:46	&	2.21	&	IR/Opt	&	6	&	10.49	&	43.39	\\
8482	&	GOODS-N	&	12:37:07	&	+62:14:08	&	2.49	&	X-ray/Opt	&	83	&	11.24	&	42.22	\\
6976	&	GOODS-N	&	12:37:23	&	+62:13:04	&	1.59	&	X-ray/Opt	&	2	&	10.59	&	42.92	\\
10421	&	GOODS-N	&	12:37:23	&	+62:15:38	&	2.24	&	X-ray/IR/Opt	&	11	&	10.57	&	42.94	\\
30014	&	AEGIS	&	14:19:21	&	+52:51:36	&	2.29	&	X-ray/Opt	&	4	&	10.56	&	41.51	\\
31250	&	AEGIS	&	14:18:33	&	+52:43:13	&	2.14	&	IR	&	129	&	10.14	&	42.38	\\
17754	&	AEGIS	&	14:19:35	&	+52:51:09	&	2.3	&	X-ray	&	45	&	11.35	&	42.30	\\
17664	&	AEGIS	&	14:19:38	&	+52:51:38	&	2.19	&	X-ray	&	26	&	10.93	&	42.64

\enddata
\label{table1}
\end{deluxetable*}

\subsection{Stellar Masses and Star Formation Rates}\label{sed}
We estimate the stellar masses and SFR of our sources by fitting their spectral energy distributions \citep[SED, full details are described in][]{aza16}.
We use multi-wavelength photometric data from the 3D-{\it HST} catalogs \citep{skel14} and fit the data with the FAST SED fitting code \citep{kri09}, adopting the FSPS stellar population synthesis model \citep{con09} and the \citet{cha03} initial mass function.
The star formation history is parametrized by a delayed exponentially declining model, where $\mathrm{SFR}(t) \propto te^{-t/\tau}$, $\tau$ being the characteristic star formation timescale.
Dust attenuation is described by the \citet{cal00} attenuation curve.

The SEDs of AGN may include a red power law in the MIR and/or a blue power law in the UV \citep[e.g.][]{kor99,don12}.
The majority of the AGN in our sample are type 2 (only six have a broad component), such that the optical light is dominated by the host galaxy.
To account for the potential contribution to the SED from the AGN, we subtract a UV and/or an IR power law from the observed photometric data, as described in \citet{aza16}.
Power law templates spanning a range of slopes and normalizations are created and subtracted from the observed photometric data before fitting the power-law subtracted photometry with FAST.  We accept the fit with the smallest $\chi^2$ as the best fit.
We create a blue power law at rest-frame wavelengths $<1$ \um ~and a red power law at rest-frame wavelengths $>1$ \um.
The normalization of the blue and red power laws can vary from 0 to 1 times of the fluxes in the U and IRAC channel 3 (5.8\um) bands, respectively.
The spectral indices of the blue and red power laws can vary from 0 to 0.5 and -5 to -0.5, respectively \citep[slightly modified from values in ][]{ive02, don12}.
The power law subtraction results in an average 0.13 dex decrease in the SFR and an average 0.03 dex decrease in the stellar mass of the AGN host galaxies.

\section{Outflow Detection and Analysis}\label{results}
In this section we report the methods used to detect outflows and their properties, including their kinematics, 
optical line ratios, and physical extent in the MOSDEF data.

\begin{figure*}[p]
	\centering
		\includegraphics[width=\textwidth]{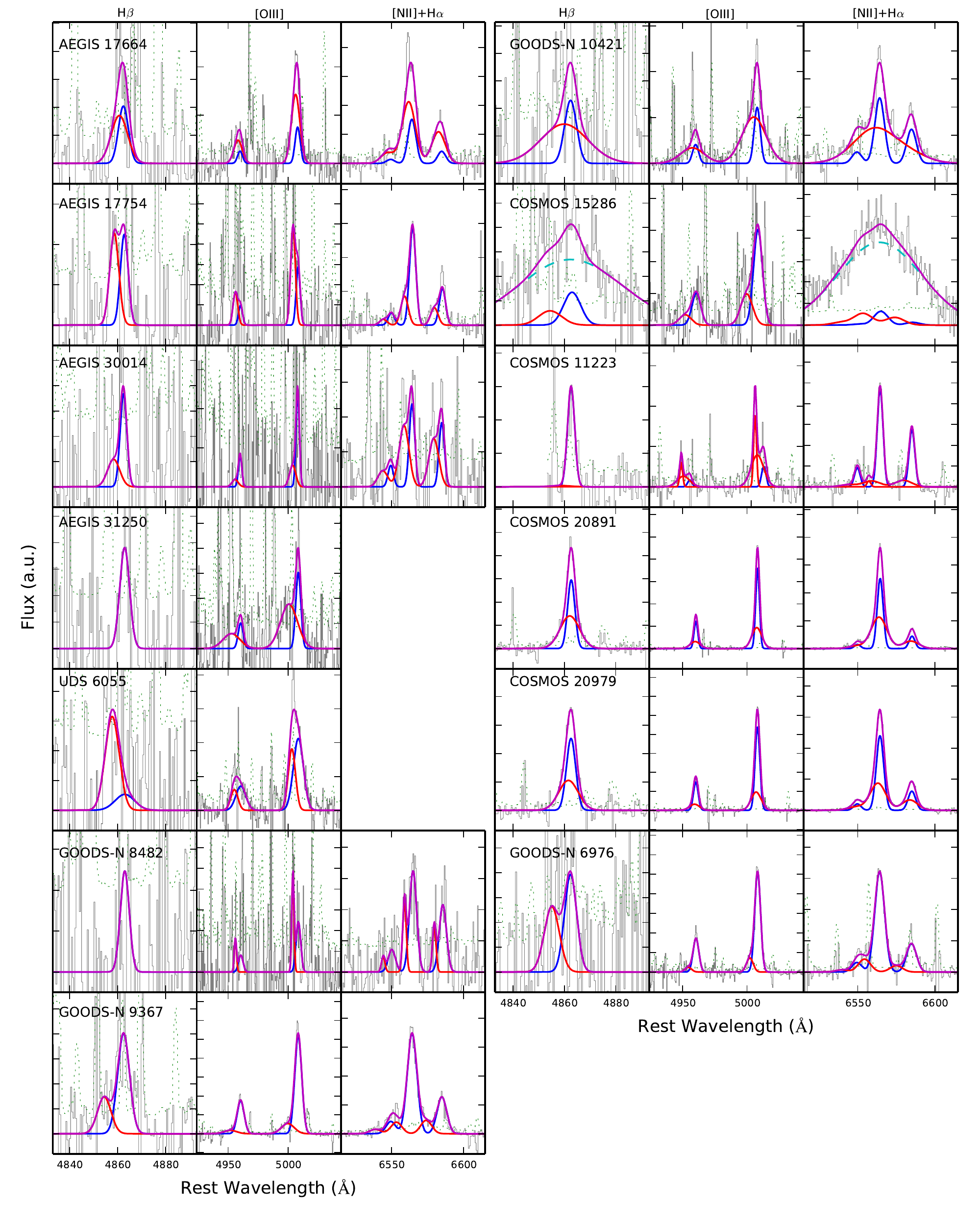}
		\caption{1D spectra showing the \hbeta ~(left), [OIII] (middle), and \halpha ~and [NII] (right) emission lines of all 13 AGN outflows not identified as potential mergers. The thin gray lines show the observed spectra; the solid blue curves show the best-fit narrow-line components; the dashed cyan curves show the best-fit broad-line components; the red curves show the best-fit outflow components; and the thick magenta curves show the best-fit total emission line profiles. Each panel is scaled to arbitrary units.}
	\label{fig:1D}
\end{figure*}

\subsection{Detection of Outflows in AGN} \label{agnoflw}

\begin{figure*}[!th]
	\centering
		\includegraphics[width=\textwidth]{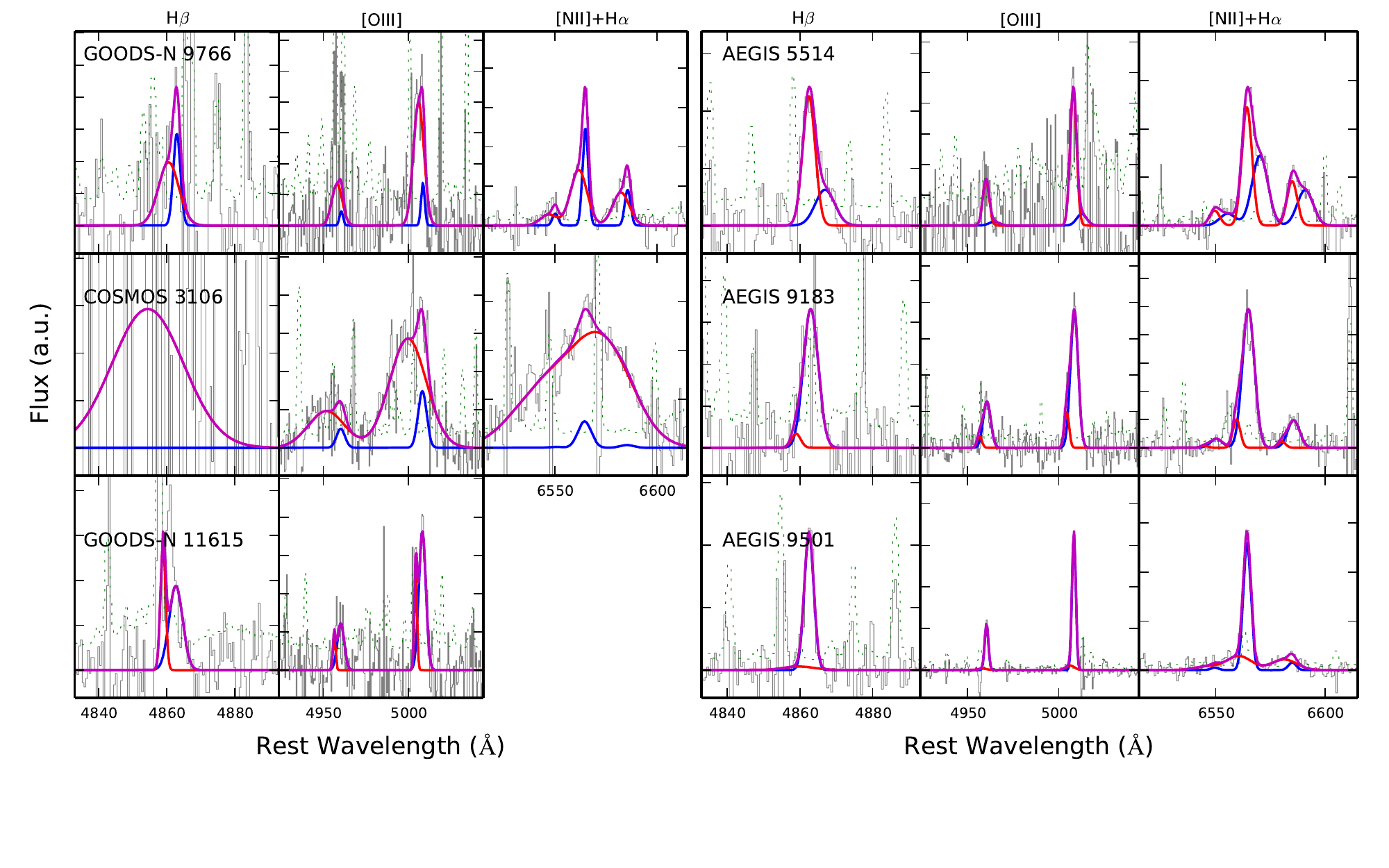}
		\caption{Same as Figure \ref{fig:1D} but for outflow candidates identified as potential mergers.}
	\label{fig:1D_mergers}
\end{figure*}

First, we identify AGN with {\it potential} outflows in our sample.
For 25 of the 67 AGN in our sample, the best fit emission line model with an outflow component is preferred, with the null hypothesis of a single component rejected at a $p$-value less than 0.01.
Of the six AGN with a broad-line component, we detect a significant outflow component in only one source, COSMOS 15286.

To reduce the chance of a false detection potentially due to overfitting noise in the spectra, we further tighten our outflow selection criteria by selecting only those AGN satisfying the following conditions:
\begin{enumerate}
\item S/N $>$ 3 for both the narrow-line and outflow components for at least one of the \hbeta , [OIII]$\lambda 5008$, [NII]$\lambda 6585$ and/or \halpha ~emission lines;
\item a velocity shift between the outflow and narrow-line components of $>$ 20 \kmps , roughly corresponding to the spectral resolution.
\end{enumerate}
The first criterion is required because of the larger number of sky lines in our observed NIR wavelengths, so it is 
common for one or more sky lines to affect these emission lines.
A total of 19 outflow candidates satisfy these more stringent criteria; their 1D spectra are shown in Figures \ref{fig:1D} and 
\ref{fig:1D_mergers}, with the best-fit models overlaid.

On-going galaxy merger events can disturb the gas kinematics in galaxies and could result in a second kinematic component in the spectra that appear similar to that from an outflow.
While it is entirely possible that outflows can happen during a merger phase, kinematically-disturbed gas from the merger event can contaminate the measurement of the kinematics of the outflows in mergers.  As our aim is to use the MOSDEF
spectra to characterize the outflowing gas, we remove sources which are potentially undergoing merger events from our sample, in order to avoid potential contamination from disturbed gas that is not outflowing.  
The study of outflows in galaxy mergers is better served using integral field spectroscopy, which is beyond the scope of this paper.

We identify potential mergers in the sample using the {\it HST} images of each source observed 
in the F160W and F606W bands.
Source features suggesting the presence of double nuclei are considered potential mergers.
We consider a source to have indication of double nuclei when two distinct peaks in brightness are separated by less than $\sim 1''$ (corresponding to $\sim 8$ kpc at $z\sim 2$).

\begin{figure}[!ht]
	\centering
		\includegraphics[width=0.5\textwidth]{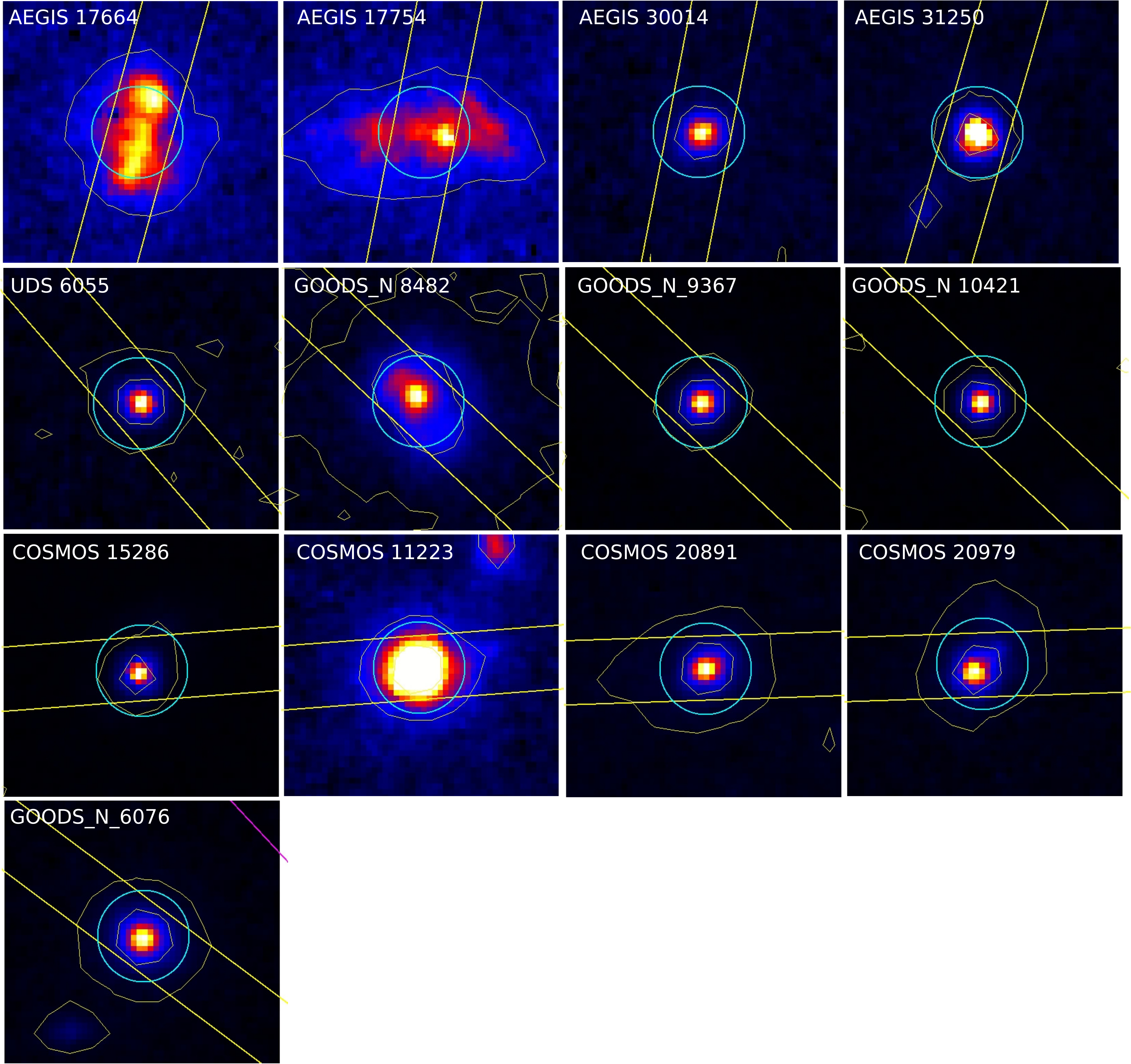}
		\caption{{\it HST} F160W band images of all 13 AGN outflows not identified as potential mergers. The MOSFIRE slit is shown in yellow lines, and a green circle of radius 0.5' centered at the source position is shown for reference. Note that the F606W band images, which are not shown here, are also inspected.}
	\label{fig:HST}
\end{figure}

\begin{figure}[!ht]
	\centering
		\includegraphics[width=0.5\textwidth]{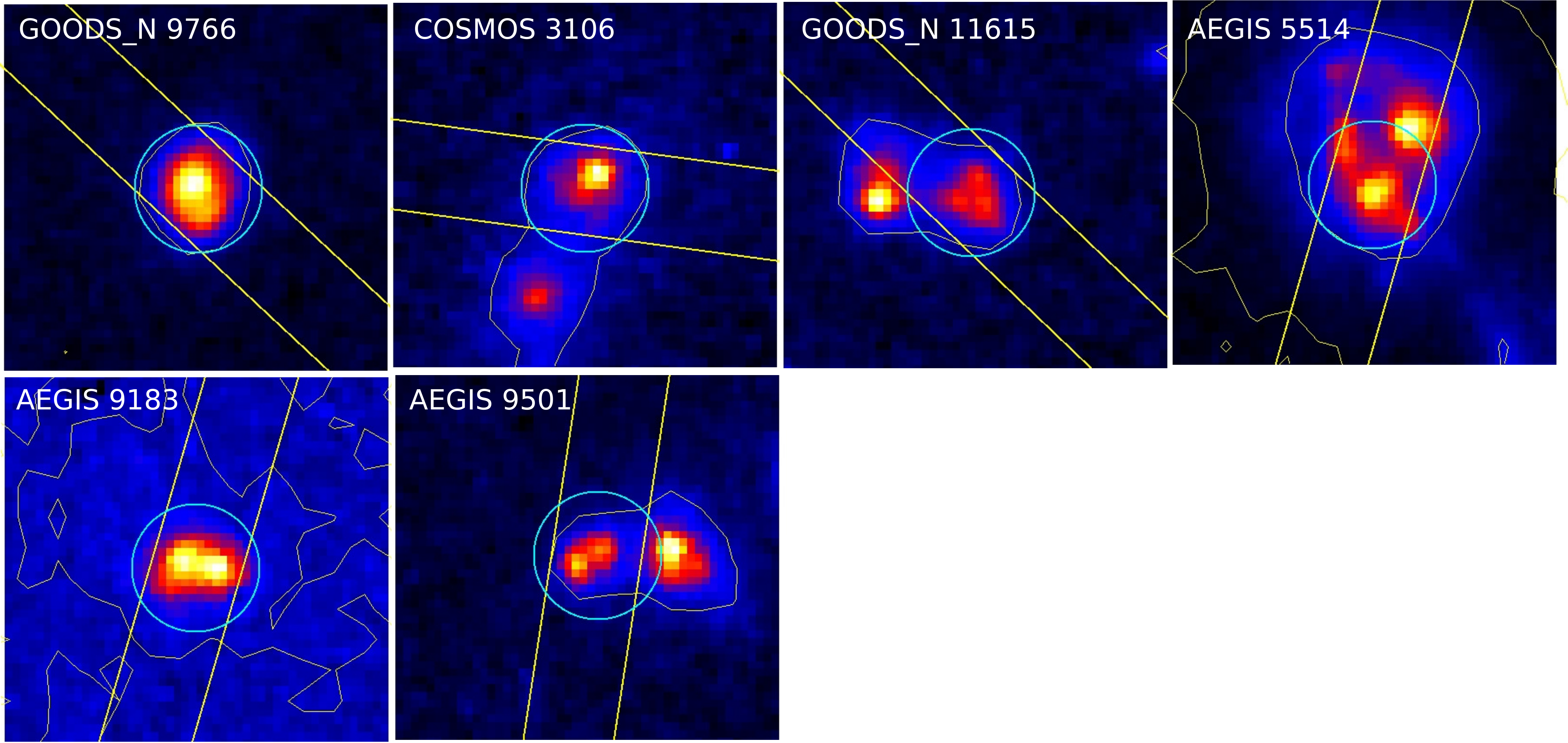}
		\caption{Same as Figure \ref{fig:HST} but for AGN outflow candidates identified as potential mergers.}
	\label{fig:HST_mergers}
\end{figure}

13 of the 19 outflow candidates are classified as non-mergers, and their {\it HST} images in the F160W band are shown in Figure \ref{fig:HST}.
Most of these outflow candidates exhibit a round morphology, without multiple distinct peaks in brightness.
The morphology of AEGIS 17664 is highly elongated with some brightness variation but is not separable into multiple distinct peaks.
This specific morphology is likely produced by an edge-on disk, therefore we include this source in our outflow sample.
The elongated morphology of AEGIS 17754 also suggests an edge-on disk.  
A total of six sources are found to contain possible double nuclei and are excluded from our outflow sample; the F160W
{\it HST} images of these sources are shown in Figure \ref{fig:HST_mergers}.
Although the F160W image of GOODS-N 9766 shows a point-like morphology, its F606W image clearly shows a ring morphology \citep[see][]{coil15}, so it is considered a potential merger.
After excluding potential mergers, there are 13 AGN with identified outflows, out of a total of 67 AGN, corresponding to $19\%$ of the full AGN sample.

\subsection{Detection of Outflows in Galaxies} 
As all MOSDEF targets with a reliable redshift have been fit using the same emission line fitting procedure described above to identify optical AGN, we can compare the incidence of outflows detected in the emission line spectra of AGN and 
galaxies in our sample.
A significant outflow component satisfying the criteria detailed in Section \ref{agnoflw} is detected in 21 
out of the 457 galaxies used in this study.
After excluding potential mergers upon inspection of their {\it HST} images, there are a total of 8 galaxies with outflows detected in emission, corresponding to $1.8 \%$ of the total galaxy sample.
These sources could potentially be AGN which are missed by our identification. 
In particular, two of the galaxies with outflows do not have coverage of \halpha ~and [NII], preventing us from performing optical AGN identification.
With an incidence rate of $19\%$, outflows detected in emission are therefore identified in MOSDEF at a rate $\sim$10 times higher for galaxies with a 
detected AGN than for inactive galaxies.  
While outflows are commonly detected in absorption in star-forming galaxies at similar redshifts \citep[e.g.][]{stei10}, our results here provide a direct comparison of the detection rate of outflows in emission in AGN and inactive galaxies.
This factor of 10 difference indicates that the outflows are likely AGN-driven (we return to this point in Section \ref{energetics}).
A detailed study of outflows detected in emission in MOSDEF galaxies is presented in \citet{free17}.
The slight difference in the number of galaxy outflow detections in that paper compared to here is likely due to small differences in the fitting procedures used in the studies.

\begin{figure*}[hbtp]
	\centering
		\includegraphics[width=0.7\textwidth]{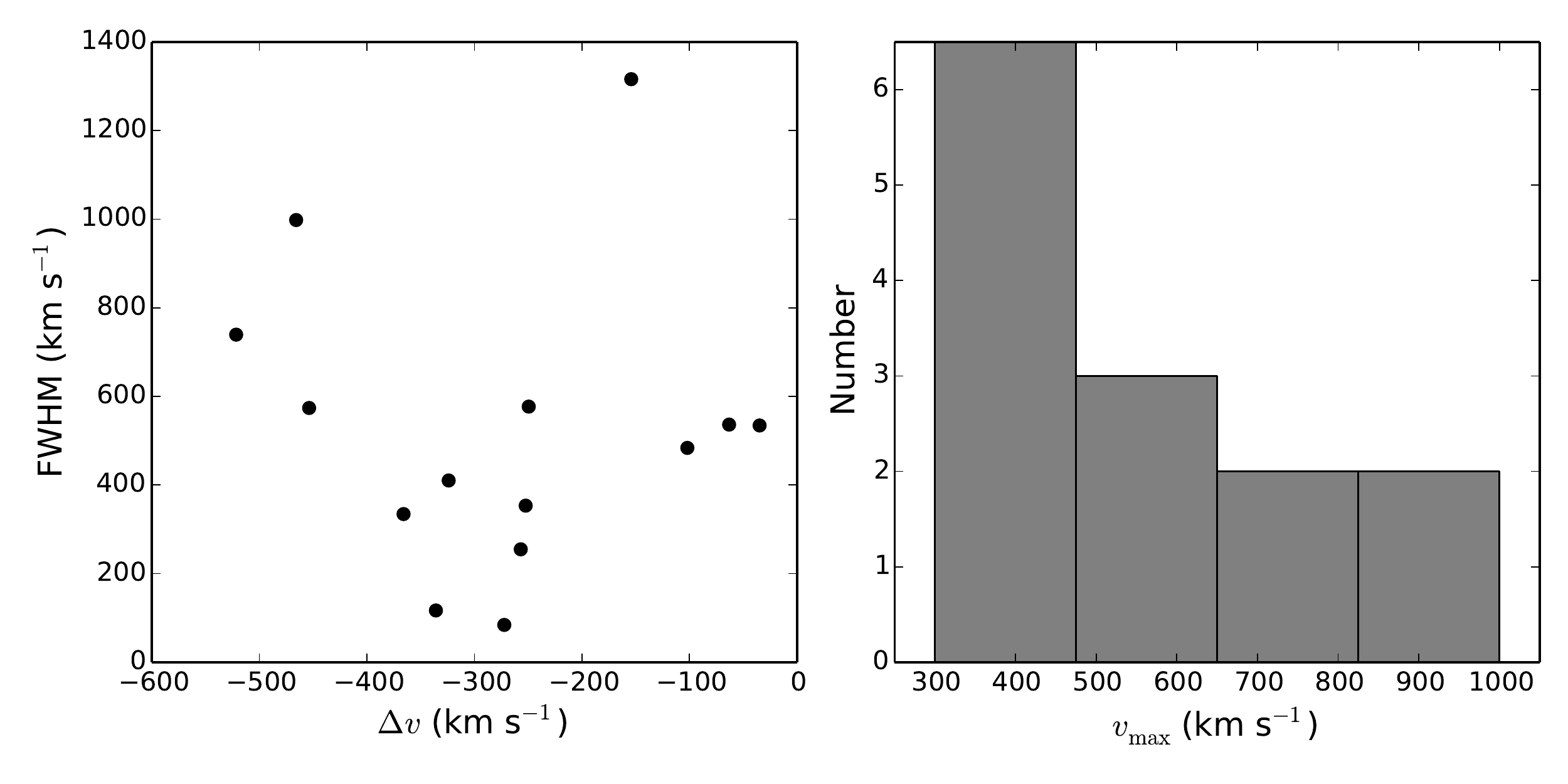}
		\caption{Left: Distribution of the velocity widths (FWHM$_v$) and velocity offsets ($\Delta v$) for the AGN outflows. Right: Distribution of the maximum velocity ($v_\mathrm{max}$) of the AGN outflows.}
	\label{fig:vhist}
\end{figure*}

\subsection{Kinematics} \label{kin}

By definition in our detection algorithm, all of the outflow components have a velocity centroid that is 
blueshifted relative to the narrow-line component (though the outflow may have some emission that is redshifted 
relative to the narrow-line emission, e.g., COSMOS 10421).
The velocity offset between the centroids of the narrow-line and outflow components $\Delta v$ ranges from $\sim-30$ to $-500$ \kmps , with a median of $\sim-270$ \kmps , while the FWHM of the outflow components ranges from $\sim$100 to 1300 \kmps , with a median of $\sim500$ \kmps .
The FWHM values reported are deconvolved with the instrumental resolution, as determined from sky lines (see Section \ref{opt}).
Figure \ref{fig:vhist} (left panel) shows the distribution of FWHM versus $\Delta v$.
In general, outflows with high velocity shifts ($\Delta v$) have relatively high FWHM, though outflows with high FWHM do not necessarily have high velocity shifts.  Similar findings in SDSS AGN are also reported in \citet{woo16}.

The velocity shift $\Delta v$, which shows the average projected line-of-sight velocity of the outflowing gas, depends on the geometry of the outflow and the extinction by dust in the galaxy, and does not directly reflect the average bulk velocity of the outflowing gas.
The average velocity of the outflow is better represented by the maximum projected velocity observed, corresponding to the most blueshifted part of the emission line profile.  As the S/N of the spectrum at this location may be low, we 
estimate this maximum velocity by $v_\mathrm{max}=|\Delta v| + \mathrm{FWHM}/2$ \citep[e.g.][]{rup05, har12}.
Figure \ref{fig:vhist} (right panel) shows the distribution of maximum velocity in our sample, which 
ranges from $\sim$ 300 to 1000 \kmps , with a median of $\sim 540$ \kmps.

We have also considered other kinematic measures such as the non-parametric line width containing $80\%$ of the flux \citep[$w_{80}$, see e.g.][]{zak14, har16}.
However, this measure depends on the flux ratio between the narrow-line and outflow components, which can vary among different emission lines, giving different results for different lines.
While studies that use this measure often have only one emission line available, we simultaneously fit the [OIII], \halpha , [NII] and \hbeta ~emission lines in this study.
For our dataset, the kinematic measures from our simultaneous fit results, utilizing information in all the emission lines, provide a more consistent and better constrained measurement of the gas kinematics.
For completeness, we compare the ratio of $w_{80}$ in [OIII] to $v_\mathrm{max}$ in our outflow sample.
Using our simultaneous fit results to calculate $w_{80}$ in [OIII], it is very consistent with $v_\mathrm{max}$, having a ratio of $1.05 \pm 0.28$.
If we calculate $w_{80}$ in [OIII] by fitting the [OIII] line alone, the ratio slightly increases to $1.35 \pm 0.47$, but is still consistent with unity given the variance.

\subsection{Line Ratios} 

\begin{figure*}[htbp]
	\centering
		\includegraphics[width=\textwidth]{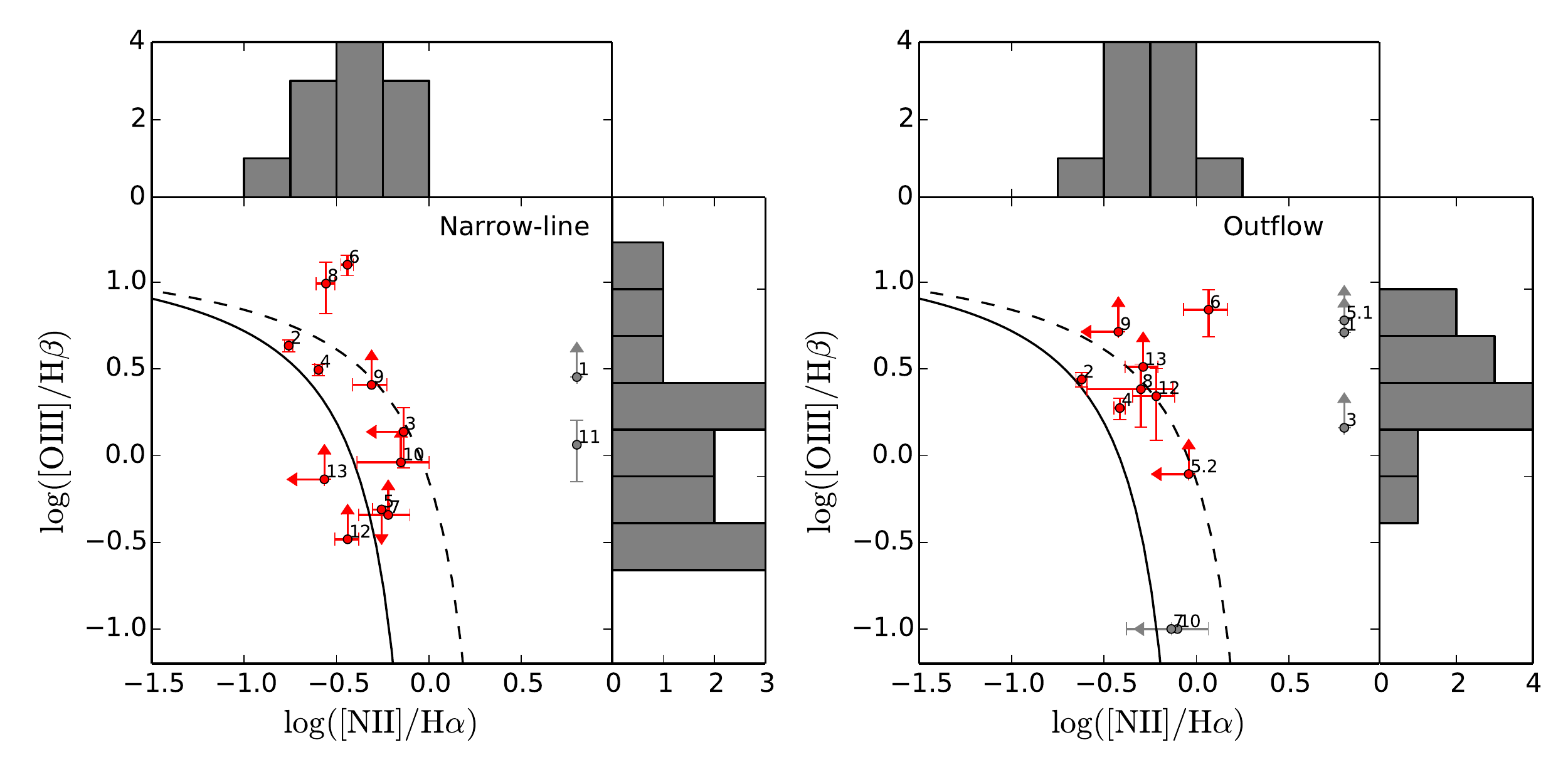}
		\caption{[NII] BPT diagrams for gas in the narrow-line component (left) and outflow component (right) of the AGN with outflows in our sample. The solid lines show the \citet{kau03} demarcation line, while the dashed line show the \citet{kew13} demarcation line at $z=2.3$. The gray points show sources where only one line ratio is measured. Upper and lower limits are also included in the histograms. The outflow component line ratios are generally shifted towards the AGN region compared with the narrow line component. The sources in each panel are labelled with the same numbers, which correspond to the row numbers in Table \ref{table1}.}
	\label{fig:BPT}
\end{figure*}

The unique dataset obtained by the MOSDEF survey, which provides simultaneous coverage of all of 
the required optical emission lines at $z \sim 2$, allows us to probe the excitation mechanism of the 
narrow-line and outflow emission with the optical diagnostic BPT diagram \citep{bal81, vei87}.
The excitation mechanism of the individual gas components allows us to test the picture of the quenching of star formation by AGN-driven outflows.
If there is increased excitation by star formation in the outflowing gas component, it will argue against the idea that the outflows are AGN-driven and these outflows quench star formation. 
Figure \ref{fig:BPT} shows the [NII] BPT diagrams for the narrow-line (left) and outflow (right) components of 
each of our AGN with outflows.
The \halpha ~and \hbeta ~fluxes of the narrow-line emission are corrected for Balmer absorption in the galaxy.
Sources with S/N$<$3 in one or both of the line ratios are shown with $3\sigma$  limits. 
Also shown are histograms for $\log(\mathrm{[OIII]}/\mathrm{H}\beta)$ and $\log(\mathrm{[NII]}/\mathrm{H}\alpha)$, which include sources where only one of the line ratios is significant.

Compared with the narrow-line components, the line ratios for the outflow components are shifted towards 
the AGN region in this diagram.  All of the outflow component line ratios lie above the \citet{kau03} line, indicating 
a likely contribution from AGN in their excitation.
For the narrow-line ratios, only two sources lie above the maximum starburst line of \citet{kew13}, in contrast to four
sources using the outflow components. 
This indicates that the excitation of the outflow emission is dominated by AGN rather than star formation for the bulk of the AGN outflows.  
The outflows are unlikely excited by shocks alone, but it is possible to explain the observed line ratios with a combination of shocks and stellar radiation \citep[e.g.][]{kew13a,new14}.
Given that these sources have known AGN, it is most likely that the gas is photoionized by the AGN.
It is worth noting that many of the narrow line components are in the ``composite'' region between the \citet{kau03} and \citet{kew13} lines.
We will return to the BPT diagram in Section \ref{pos} to discuss the lack of positive AGN feedback observed.

\subsection{Physical Extent} \label{ext}

\begin{figure*}[htbp]
	\centering
		\includegraphics[width=0.7\textwidth]{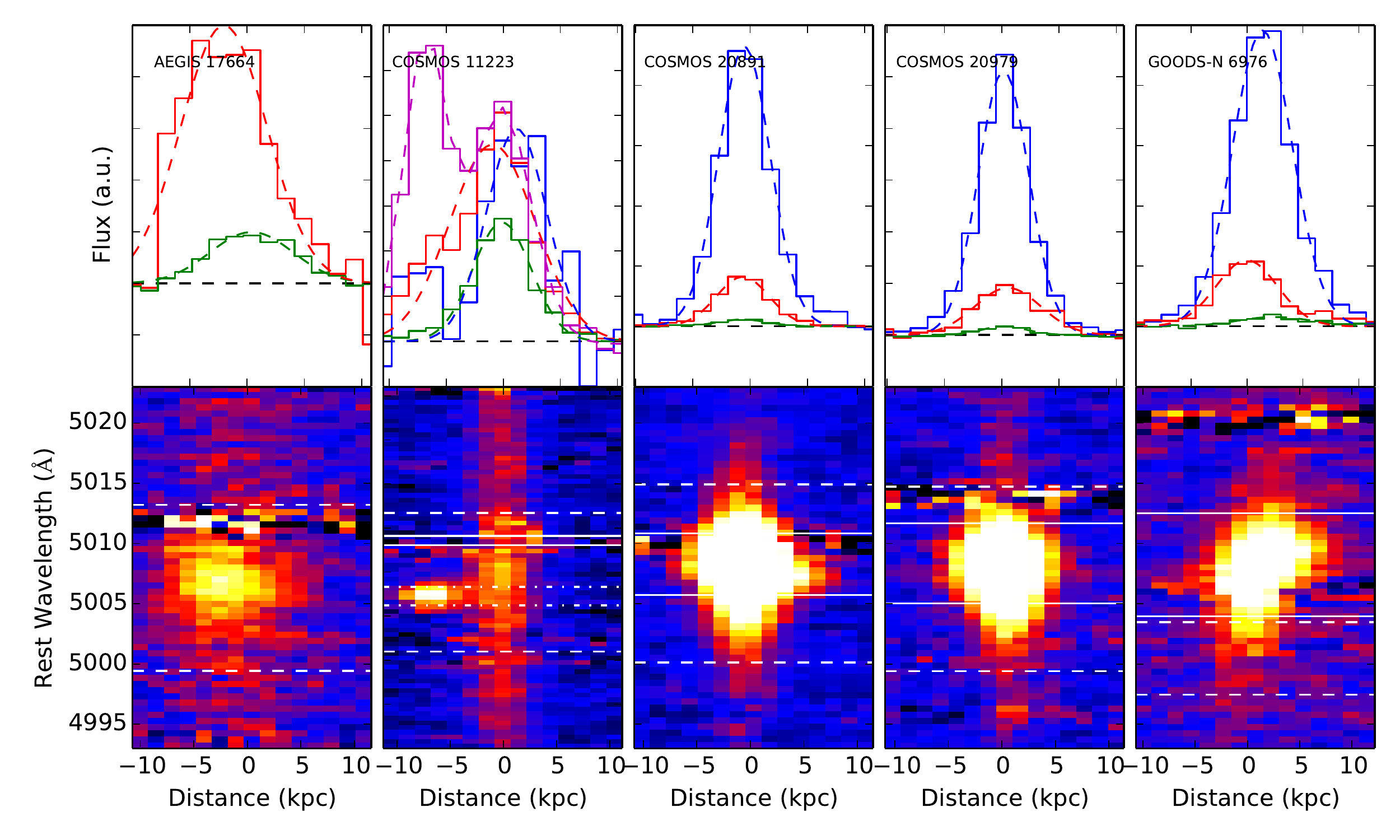}
		\caption{Spatial profiles (upper) derived from 2D spectra (lower) of the [OIII]$\lambda 5008$ emission line for AGN outflows with significant spatial extent detected in this emission line. In the 2D spectra the solid white lines show the wavelength range used to extract the narrow-line spatial profile, while the dashed white lines show the wavelength range used to extract the outflow spatial profile. For COSMOS 11223, the wavelength range dominated by the additional, faster outflow component is shown with dashed-dotted white lines. If two wavelength ranges overlap, the overlapped region is excluded from the wider wavelength range during the extraction. (See text for details.) In the spatial profiles the narrow-line, outflow, and continuum profiles are shown in blue, red and green, respectively; the second, faster outflow component of COSMOS 11223 is shown in magenta.  Dashed curves show the best-fit Gaussian function to each spatial profile.}
	\label{fig:spatial_oiii}
\end{figure*}

\begin{figure*}[thbp]
	\centering
		\includegraphics[width=0.95\textwidth]{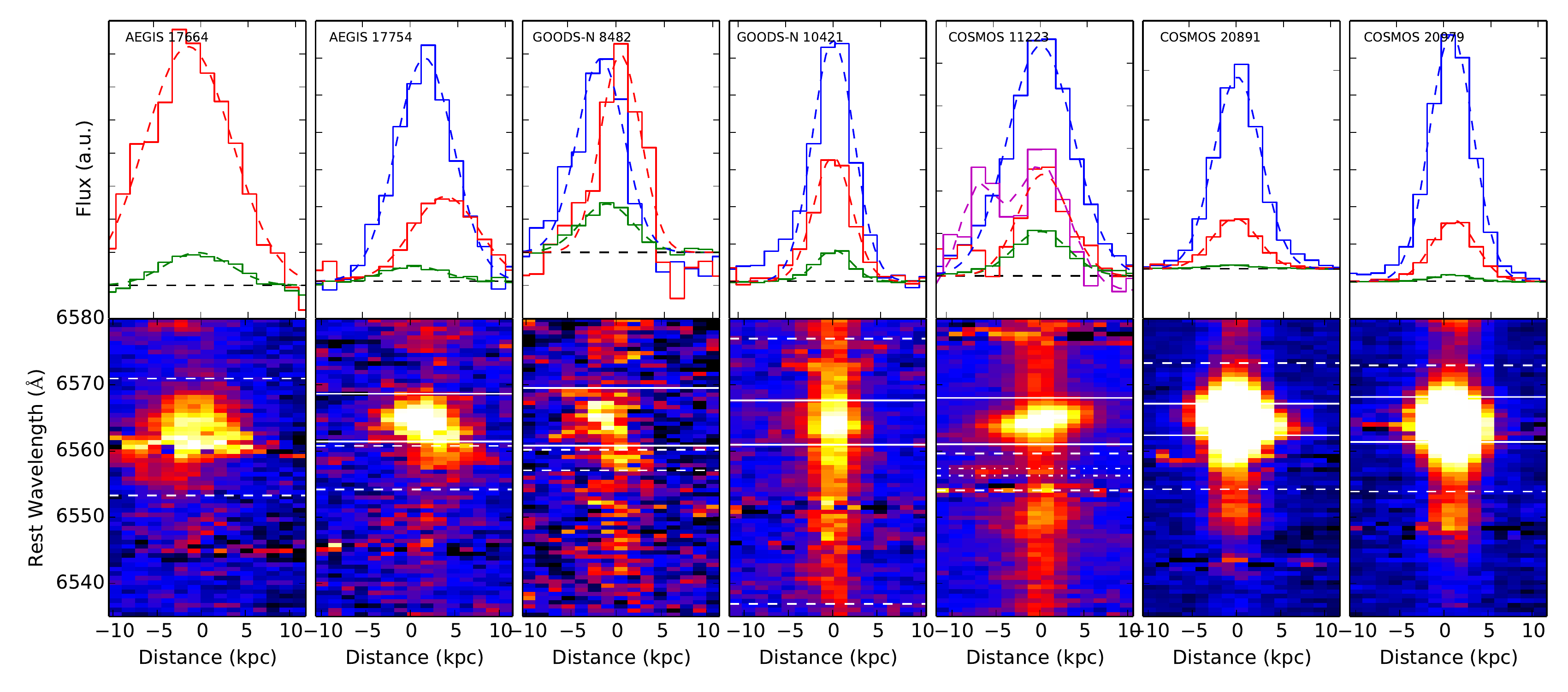}
		\caption{Same as Figure \ref{fig:spatial_oiii} but for the \halpha ~emission line.}
	\label{fig:spatial_ha}
\end{figure*}

One of the key properties of AGN-driven outflows is their physical extent, which is crucial information needed to determine the impact outflows have on their host galaxies and whether they can successfully expel gas out of the halo of the galaxy and quench star formation.
With the slit spectra from MOSDEF, we have spatial information along the slit direction, allowing us to determine the physical extent of the outflows in one spatial direction.
To investigate the potential presence of physically-extended emission, we create spatial profiles from the 2D spectra for the narrow-line and outflow components for the [OIII]$\lambda 5008$ 
and \halpha ~emission lines, as well as for the continuum.
We first select narrow-line or outflow dominated wavelength ranges, over which the narrow-line or outflow flux from the 1D Gaussian fit is higher than the sum of all of the other components.
We also limit the wavelength range to be within 2$\sigma$ of the central wavelength of the respective component. 

For AEGIS 17664, the narrow-line component is weaker than the outflow component in all wavelengths, so only an outflow wavelength range is selected according to the criteria above.
For sources where the narrow-line wavelength range is enclosed within the outflow wavelength range, such as COSMOS 20891, the narrow-line wavelength range is excluded from the outflow wavelength range according to the criteria above, since it is weaker than the outflow flux.
The continuum is selected in wavelength ranges far from emission lines.
We then sum the fluxes in the 2D spectra along the wavelength axis within the specific wavelength range to create spatial profiles for each component: narrow-line, outflow, and continuum.

To quantify the physical extent of the emission in each component, we fit a single Gaussian function to each spatial profile.
The difference between the Gaussian centroids of the narrow-line and outflow components, $|\Delta x|$, will give the projected spatial offset between the narrow-line region and the outflows.
The width of each Gaussian is deconvolved from the seeing, measured from the star observed on every MOSDEF mask.
An outflow is resolved if the deconvolved width of the outflow component is significantly non-zero (more than three times its uncertainty).
We combine these two measurements to estimate the full physical extent of the outflow from the center of the narrow-line region.
We define $r_{10}$ as $|\Delta x| + \mathrm{FWTM}/2$, where FWTM is the full width at 1/10 of the maximum of the outflow component.

The spatial profile measurements are shown in Table \ref{table2}.
Six sources have their outflow component spatially offset from the narrow-line component in at least one of the emission lines while seven are resolved.
Eight sources have a significantly non-zero $r_{10}$, ranging from 2.5 to 11.0 kpc.
Figures \ref{fig:spatial_oiii} and \ref{fig:spatial_ha} show the 2D spectra and spatial profiles of the [OIII]$\lambda 5008$ and \halpha ~emission lines for the sources with significantly non-zero $r_{10}$.
We measure $r_{10}$ for both the [OIII] and \halpha ~emission lines because often one of the lines is impacted by a sky line.
In the cases where $r_{10}$ is well measured in both lines, the spatial extent in the two lines generally agrees, with the [OIII] extent ranging from $\sim 0.7-1.5$  that in \halpha.

\subsection{COSMOS 11223}

\begin{figure*}[htbp]
	\centering
		\includegraphics[width=0.8\textwidth]{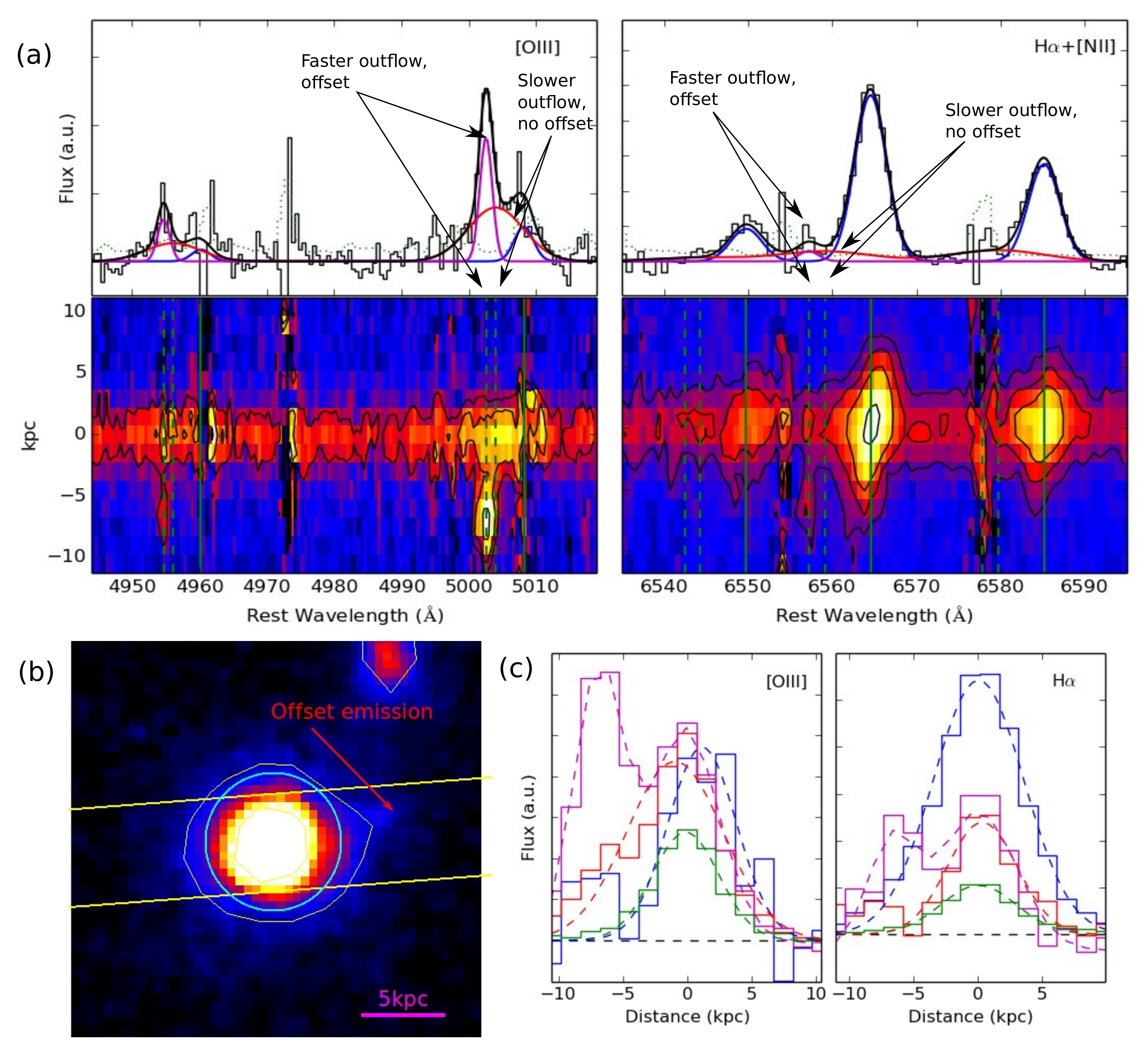}
		\caption{(a) 1D (top) and 2D (bottom) spectra of COSMOS 11223. The 1D spectrum is extracted from the central 13 spatial pixels. The thin black line shows the observed spectra; the blue curve shows the best-fit narrow-line component; the red curves shows the best-fit for the slower outflow; the red curve shows the best-fit for the faster outflow; and the thick black curve shows the best-fit total emission line profile. (b) {\it HST} F160W band images of COSMOS 11223. Note the nearby source at the top-right corner at a similar projected position with the offset emission. The MOSFIRE slit is shown in yellow lines, and a green circle of radius 0.5' centered at the source position is shown for reference. (c) Spatial profiles for the [OIII]$\lambda 5008$ (left) and \halpha ~(right) emission lines. 
Two outflow components at different velocities are detected in this source. The faster outflow is spatially offset from the narrow-line region by 6.8 kpc while the slower outflow is not.  The spatially offset outflow is seen in the {\it HST} image as a faint emission along the MOSDEF slit.}
	\label{fig:11223}
\end{figure*}

Among the AGN with outflows, COSMOS 11223 is particularly interesting, as this source has {\it two} significant outflow components at different velocities.  It is the only source in our sample where the best fit requires more than one outflow component.  Figure \ref{fig:11223} shows the 1D and 2D spectra of this source, along with the {\it HST} F160W image and the spatial profile. 

The slower outflow has a velocity offset of -253 \kmps and a FWHM of 580 \kmps.  It is detected in \halpha ~(S/N=3) and at high S/N in the [OIII] lines (S/N=5 for both lines combined). [NII] has a marginal S/N of 2.6. 
The faster outflow has a velocity offset of -340 \kmps and a FWHM of 136 \kmps.  It is detected at very high S/N in the [OIII] lines (S/N=13 for both lines combined) and can be seen clearly in each of the [OIII] lines.  
The detection at two wavelengths indicates that it is not a chance projection of a single emission line from a source at another redshift.  This outflow is not detected at [NII], as there is a sky line at this wavelength, but it is detected at \halpha ~(S/N=5).  
On the BPT diagram, the slower outflow is not detected in either \hbeta ~or [NII], such that it can not be placed on the BPT diagram. 
The faster outflow is detected in one of the lines in both line ratios, resulting in a limit that lies just below the \citet{kew13} demarcation line, and is labelled ``14'' in Figure \ref{fig:BPT}.

In terms of spatial extent, the slower outflow is not significantly resolved or offset from the narrow-line region in this source, but the faster outflow is both resolved and offset.  It has the largest physical extent of any outflow in our sample, $r_{10}$=11.0 kpc in [OIII].  The {\it HST} F160W, F814W, and F606W images all show faint extended emission in the direction of the outflow as detected on the MOSDEF slit.  The {\it HST} images show a faint nearby source (COSMOS 10856) located $1.6''$ away from COSMOS 11223 (see Figure \ref{fig:11223}). 
This nearby source is outside of the MOSDEF slit, but its projected position along the MOSDEF slit is somewhat similar to the position of the spatially extended outflow component.
In the 3D-{\it HST} catalog \citep{mom15, bram12}, this nearby source has a grism redshift estimate of $z=0.52-0.63$ and a photometric redshift estimate of $z=0.81-0.90$, within the $95\%$ confidence intervals.
There is no current spectroscopic redshift available for this source, such that we can not strictly rule out the possibility of a tidal interaction between this source and COSMOS 11223, if they are at the same redshift.  We are in the process of conducting Keck/OSIRIS integral field spectroscopy follow-up of this source, to better
determine the nature of this very extended outflow.

\capstartfalse
\begin{deluxetable*}{lrrrrrrrrr}
\centering
\tablecaption{Kinematic and Spatial Measurements}
\tablehead{
\colhead{ID}&\colhead{$\Delta v$}&\colhead{FWHM}&
\colhead{$v_\mathrm{max}$}&\colhead{Offset ([OIII])}&
\colhead{FWHM ([OIII])}&\colhead{$r_{10}$ ([OIII])}&
\colhead{Offset (\halpha)}&\colhead{FWHM (\halpha)}&\colhead{$r_{10}$ (\halpha)}
\\ \colhead{} & \colhead{(\kmps)} & \colhead{(\kmps)} & \colhead{(\kmps)}
 & \colhead{(kpc)} & \colhead{(kpc)} & \colhead{(kpc)} & \colhead{(kpc)}
  & \colhead{(kpc)} & \colhead{(kpc)}
}
\startdata
6055	&	-324	&	410	&	529	&	0.3	$\pm$	0.4	&	\tablenotemark{a}			&	0.3	$\pm$	0.4	&	\tablenotemark{b}			&	\tablenotemark{b}			&	\tablenotemark{b}			\\
20891	&	-35	&	534	&	302	&	0.2	$\pm$	0.1	&	{\bf 2.8	$\pm$	0.2}\tablenotemark{c}	&	{\bf 2.7	$\pm$	0.2}	&	{\bf 0.3	$\pm$	0.1}	&	{\bf 4.0	$\pm$	0.2}	&	{\bf 3.9	$\pm$	0.2}	\\
15286	&	-454	&	574	&	741	&	0.1	$\pm$	0.1	&	1.6	$\pm$	0.7	&	1.6	$\pm$	0.7	&	\tablenotemark{b}			&	\tablenotemark{b}			&	\tablenotemark{b}			\\
20979	&	-63	&	536	&	331	&	0.4	$\pm$	0.3	&	{\bf 4.8	$\pm$	0.9}	&	{\bf 4.8	$\pm$	0.9}	&	0.2	$\pm$	0.1	&	{\bf 3.6	$\pm$	0.1}	&	{\bf 3.5	$\pm$	0.1}	\\
11223a	&	-250	&	577	&	538	&	2.1	$\pm$	0.7	&	{\bf 7.4	$\pm$	0.6}	&	{\bf 8.8	$\pm$	0.9}	&	0.3	$\pm$	0.5	&	3.6	$\pm$	2.8	&	3.5	$\pm$	2.6	\\
11223b	&	-335	&	117	&	394	&	{\bf 6.8	$\pm$	0.1}	&	{\bf 4.5	$\pm$	0.3}	&	{\bf 11.0	$\pm$	0.3}	&	{\bf 6.7	$\pm$	0.7}	&	4.5	$\pm$	2.0	&	{\bf 10.7	$\pm$	1.2}	\\
9367	&	-522	&	739	&	891	&	0.5	$\pm$	0.3	&	3.4	$\pm$	1.5	&	3.5	$\pm$	1.4	&	0.2	$\pm$	0.3	&	3.5	$\pm$	1.3	&	3.4	$\pm$	1.2	\\
8482	&	-272	&	84	&	314	&	3.5	$\pm$	3.5	&	9.2	$\pm$	12.3	&	11.8	$\pm$	11.8	&	{\bf 2.0	$\pm$	0.4}	&	2.9	$\pm$	1.3	&	{\bf 4.7	$\pm$	1.3}	\\
6976	&	-366	&	334	&	533	&	{\bf 1.4	$\pm$	0.2}	&	{\bf 4.4	$\pm$	0.9}&	{\bf 5.4	$\pm$	0.8}	&	1.5	$\pm$	0.5	&	\tablenotemark{a}			&	1.5	$\pm$	0.5	\\
10421	&	-154	&	1316	&	812	&	0.0	$\pm$	0.3	&	2.3	$\pm$	1.4	&	2.1	$\pm$	1.3	&	0.1	$\pm$	0.1	&	{\bf 2.6	$\pm$	0.5}	&	{\bf 2.5	$\pm$	0.5}	\\
30014	&	-252	&	353	&	429	&	2.1	$\pm$	1.1	&	\tablenotemark{a}			&	2.1	$\pm$	1.1	&	1.6	$\pm$	0.7	&	\tablenotemark{a}			&	1.6	$\pm$	0.7	\\
31250	&	-466	&	998	&	965	&	0.4	$\pm$	0.6	&	3.7	$\pm$	2.2	&	3.8	$\pm$	2.1	&	\tablenotemark{b}			&	\tablenotemark{b}			&	\tablenotemark{b}			\\
17754	&	-257	&	255	&	384	&	1.2	$\pm$	1.0	&	2.0	$\pm$	4.6	&	3.1	$\pm$	4.3	&	{\bf 2.1	$\pm$	0.3}	&	{\bf 5.7	$\pm$	1.0}	&	{\bf 7.3	$\pm$	1.0}	\\
17664	&	-102	&	484	&	344	&	{\bf 2.4	$\pm$	0.5}	&	{\bf 6.1	$\pm$	1.0}	&	{\bf 7.9	$\pm$	1.1}	&	{\bf 1.0	$\pm$	0.3}	&	{\bf 8.6	$\pm$	0.6}	&	{\bf 8.8	$\pm$	0.6}	

\enddata
\tablenotetext{a}{FWHM is not larger than seeing.}
\tablenotetext{b}{\halpha ~measurements are unavailable in 31250, 6055 and 15286.}
\tablenotetext{c}{Spatial measurement which are significantly non-zero are shown in bold.}
\label{table2}
\end{deluxetable*}

\section{Discussion}\label{discuss}
\subsection{Outflow incidence}

In our sample of 67 X-ray, IR, and/or optically-selected AGN at $z \sim 2$, outflows 
are identified in 13 ($19\%$) of the AGN (after removing likely on-going mergers).
In the 457 non-AGN MOSDEF galaxies, only 8 ($1.8\%$) outflows are detected.
The outflow incidence detected in the MOSDEF AGN sample is similar to that found at $z \sim 0$ by \citet{mul13} using broad [OIII] emission, who find outflow signatures in $17\%$ of $>24000$ optically-selected AGN in SDSS.
The incidence of outflows in AGN at $z \sim 2$ is at least as high as that at $z \sim 0$.
As shown below in Section \ref{s/n}, the detection rate of outflows in our sample correlates with the signal-to-noise ratio of the [OIII] emission line.  
Therefore, in general, the incidence rate found here should be treated as a lower limit on the true outflow incidence.

\citet{har16} analyze the [OIII] spectra of 54 X-ray AGN at $z \sim 1.1-1.7$, detecting a second kinematic component in 14 of the AGN, corresponding to $26 \%$ of their AGN sample.  This is comparable to the outflow incidence found in our study at $z\sim2$.
If only X-ray AGN are considered in our sample, the outflow incidence is $25 \%$, in very good agreement with \citet{har16}.
At $z \sim 2$, \citet{gen14} finds 34 outflows in 110 star-forming galaxies,  while 18 of the star-forming galaxies are confirmed AGN.
While our study first defines a sample of AGN and searches for outflows in it, \citet{gen14} searches for outflows in a sample of galaxies.
Due to the difference in the nature of the parent sample, one cannot directly compare the outflow incidence in this study with that in \citet{gen14}.
In particular, one cannot directly determine the incidence of outflows among AGN at $z \sim 2$ using the results of \citet{gen14}.

\subsection{Correlation with host galaxy properties}

AGN-driven outflows are commonly believed to affect the evolution of their host galaxies, in particular by quenching star formation in galaxies in the high mass end of the stellar mass function \citep{dm05,hop06a,hop08,deb12}.
Therefore, it is informative to study the correlation between the incidence of AGN outflows with their host galaxy properties such as stellar mass and SFR.

\begin{figure*}[!ht]
	\centering
		\includegraphics[width=0.8\textwidth]{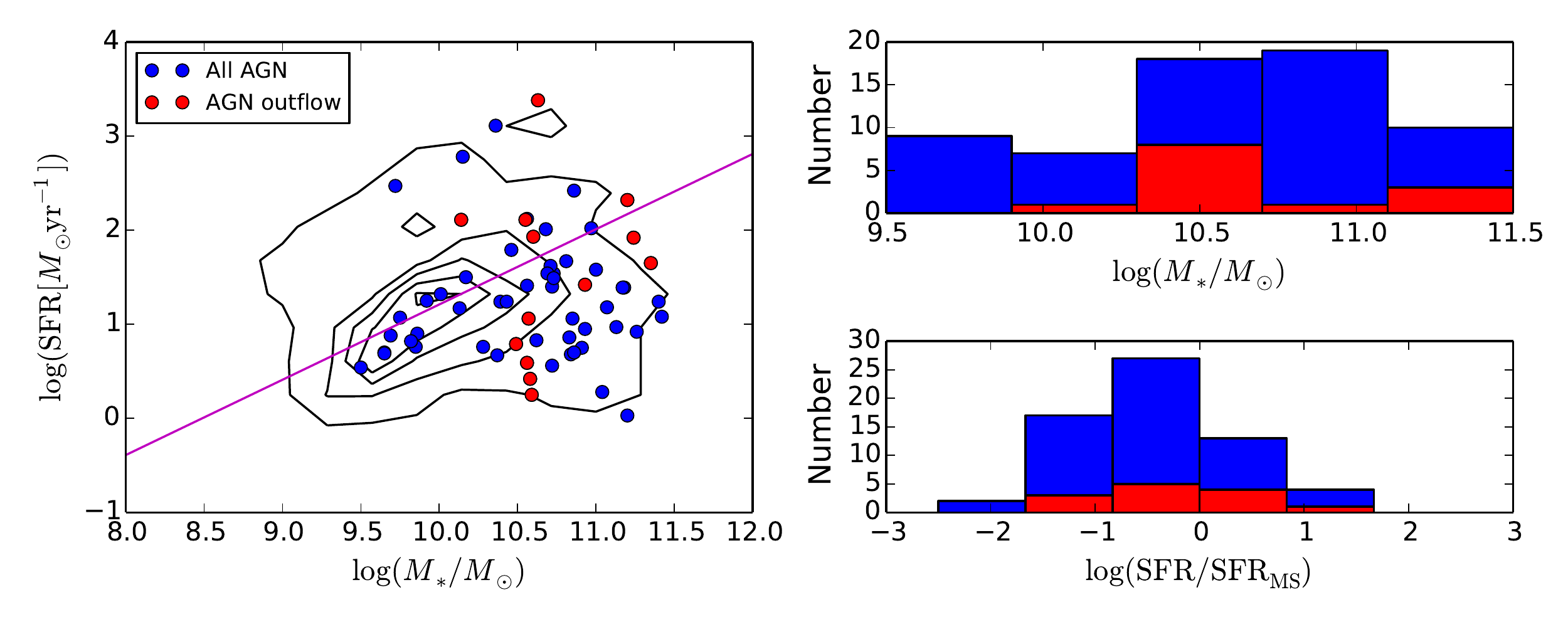}
		\caption{Left: SFR versus stellar mass for all galaxies (black contour), AGN (blue points) and AGN with outflows (red points) in the current sample. The magenta solid line shows the star formation main sequence. Upper right: Distribution of stellar mass for all AGN (blue) and AGN with outflows (red). Lower right: Distribution of $\mathrm{SFR/SFR_{MS}}$ for all AGN (blue) and AGN with outflows (red).
Outflows are detected in AGN spanning the main sequence of star formation, with no clear dependence on the stellar mass or SFR of the host galaxy.}
	\label{fig:m_sfr}
\end{figure*}

Figure \ref{fig:m_sfr} (left panel) shows the distribution of SFR versus stellar mass (see Section \ref{sed}) for all of the galaxies (shown in contours), AGN (blue filled circles), and AGN with outflows (red filled circles) in the sample used here.
The star formation main sequence for MOSDEF galaxies found by \citet{shiv15}:
\begin{equation}
\log\left(\mathrm{SFR}/M_\odot ~\mathrm{yr^{-1}}\right) = 0.8 ~\log(M_*/M_\odot) - 6.79
\end{equation}
is also shown (magenta line).
The upper right and lower right panels of Figure \ref{fig:m_sfr} 
show the distributions of stellar mass and SFR, respectively, of all AGN and AGN with outflows in our sample.
Outflows are detected in AGN spanning stellar masses of $10^{10} - 10^{11.5} \msun$.
A two-sample KS test on the distributions of stellar masses for AGN with and without outflows returns a p-value of 0.4 for the null hypothesis that the two distributions are the same.
Therefore, no significant trend of outflow incidence with host stellar mass is observed.
Similarly, the distribution of SFR with respect to the main sequence is similar for AGN with and without detected outflows.

\citet{gen14} use a sample of 44 star-forming galaxies at $z\sim2$ to report a strong stellar mass-dependent incidence of detected outflows observed in emission that increases with stellar mass at least as much as the incidence of AGN within the galaxy sample.  
They imply that the outflows are predominantly driven by AGN and that there is therefore a strong stellar mass-dependence to the incidence of AGN-driven outflows.  
However, \citet{gen14} do not first create a sample of AGN and then identify outflows within that AGN sample; they start with a sample of star-forming galaxies and identify outflows and AGN within that galaxy sample.  
This means that they cannot measure the incidence of AGN-driven outflows within the AGN population, as we do here.  

Additionally, there is a well-known selection bias that leads to AGN in higher mass galaxies being more likely to be identified, while the true incidence of AGN in galaxies is independent of stellar mass to first order \citep{aird12}.  
This implies that if outflows are more likely to be identified in AGN (as we find here), which are preferentially identified in higher mass galaxies (and seen by both \citealt{gen14} and in MOSDEF by \citealt{aza16}), then it will {\it appear} that outflows are more common in higher mass galaxies, many of which host AGN (as seen by \citealt{gen14}).  
However, it does not mean that AGN-driven outflows are actually more common in high mass galaxies, and it does not imply that there is a stellar mass dependence to the incidence of AGN-driven outflows.  
Therefore our result that the incidence of detected outflows within a sample of AGN does {\it not} show a strong mass dependence is not in conflict with the data presented in \citet{gen14}.

It is also interesting to note that we find that the incidence AGN outflows at $z\sim2$ does not appear to depend on the host SFR with respect to the main sequence.
This suggests that the incidence of AGN-driven outflows is uniform among high mass galaxies ($M_* > 10^{10} \msun$) regardless of their SFR, and could be taken to imply that the presence of AGN-driven 
outflows does not quench star formation.  
However, different methods used to estimate SFR are sensitive to star formation on different timescales, and the method of SED fitting used in this study reflects star formation over the relatively long timescale of the {\it past} $10^8$ years \citep[e.g.][]{ken98}, such that the instantaneous SFR could differ.
Therefore, the measured SFR may reflect the SFR in the past rather than the immediate effect of the outflows on the galaxies.

On the other hand, theoretical work \citep[e.g.][]{dm05,ant10} predicts that the timescale for quenching due to AGN outflows is $\sim 10^8$ yrs, such that there can be a time delay for the effect of the AGN outflow on the host galaxy to become observable on the SFR.
Moreover, AGN are known to vary significantly over a range of timescales \citep[e.g.][]{hic14, gar14}, such that the AGN may not be detectable when its effects due to feedback become observable.
Such AGN variability can also introduce scatter in any underlying correlations between
outflow properties and the AGN bolometric luminosity.

\subsection{Correlation with [OIII] luminosity and S/N}\label{s/n}

\begin{figure*}[!ht]
	\centering
		\includegraphics[width=0.65\textwidth]{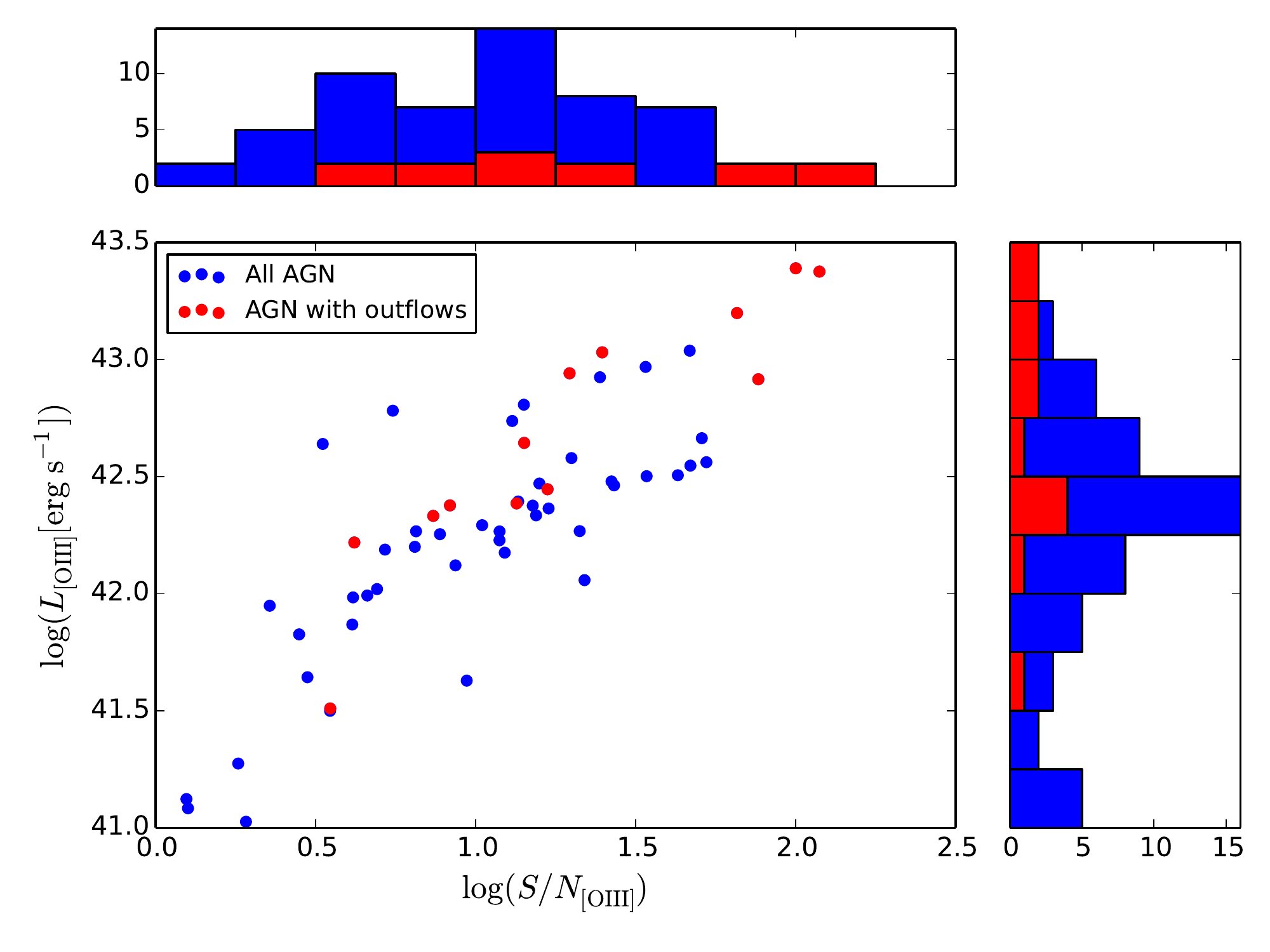}
		\caption{Distribution of $L_\mathrm{[OIII]}$ versus the signal-to-noise ratio in [OIII] for all AGN (blue) and AGN with outflows (red).
While outflows are more often found for sources with high $L_\mathrm{[OIII]}$, the same trend is also observed for the signal-to-noise of the spectra as well.}
	\label{fig:loiii}
\end{figure*}

One method that may provide hints of the driving mechanism of the outflows is to look for correlations between outflow properties and the AGN luminosity.
A common proxy for the AGN bolometric luminosity ($L_\mathrm{AGN}$) is the [OIII] luminosity ($L_\mathrm{[OIII]}$), where $L_\mathrm{AGN}$ and $L_\mathrm{[OIII]}$ are related by a constant bolometric correction factor \citep[e.g. 600 in][]{kau09}.
As we have [OIII] luminosities for all AGN in our sample, but we do not necessarily have X-ray luminosities
or IR luminosities that are dominated by light from the AGN itself (as opposed to the host galaxy), we compare the incidence of outflows with $L_\mathrm{[OIII]}$.  While  $L_\mathrm{[OIII]}$ can be affected by contamination 
from star formation in the host galaxy, \citet{aza16} estimate that on average only $\sim30$\% of the [OIII] light in AGN in the MOSDEF sample may be due to star formation.  
However, the comparison between outflow incidence and $L_\mathrm{[OIII]}$ may be complicated by the 
fact that sources with a higher $L_\mathrm{[OIII]}$ also tend to have a higher signal-to-noise ratio, which could potentially lead to an easier detection of an outflow in the spectra.
As a result, one must take into account the signal-to-noise ratio when looking for any such correlations with $L_\mathrm{[OIII]}$.

The values of $L_\mathrm{[OIII]}$ we use are corrected for dust reddening.
To determine the correction factor, we calculate the color excess from the Balmer decrement and combine this with the value of the MOSDEF dust attenuation curve at 5008\AA ~\citep{red15}.
The corrected $L_\mathrm{[OIII]}$ is on average increased by ∼0.17 dex for the AGN in our sample \citep[see][]{aza16}.
 
Figure \ref{fig:loiii} shows the distribution of $L_\mathrm{[OIII]}$ versus signal-to-noise ratio in the [OIII] line, for all AGN and for AGN with outflows.
There is a higher incidence rate of detected outflows in AGN with higher $L_\mathrm{[OIII]}$, but a similar trend is also observed in the signal-to-noise ratio.
At a given signal-to-noise ratio, outflows appear to be observed slightly more commonly in those with higher $L_\mathrm{[OIII]}$, but the difference is not significant. 
Therefore, the correlation of the incidence of AGN outflows with $L_\mathrm{[OIII]}$ cannot be tested in our sample due to its coupling with the signal-to-noise ratio.
We also compare the outflow velocity with $L_\mathrm{[OIII]}$ with no clear trend observed, which is consistent with other AGN outflow studies \citep[e.g.][]{har14}.

Our results show that the signal-to-noise ratio is an important factor in the detection of outflows, which must be taken into account when comparing outflow incidence with $L_\mathrm{[OIII]}$.
Moreover, high velocity ($\sim 1000$~ \kmps) outflows are found to occur in AGN across a range of $L_\mathrm{[OIII]}$.

\subsection{Correlation with X-ray, IR and optical AGN identification}\label{ID}
In the MOSDEF survey, AGN are identified through X-ray, IR and optical selection techniques, which are described in detail in Section \ref{data}.
It is known that different identification methods present biases in both the AGN and host properties, and such biases in the MOSDEF sample are discussed in \citet{aza16}.
In particular, they show that optical AGN identification is possibly biased towards galaxies with higher dust content, while IR AGN identification may be biased towards galaxies with lower dust content in the MOSDEF sample.
This selection bias may help constrain the driver of the outflow since it has been proposed that the radiation pressure on dust grains can lead to galaxy-wide outflows and boost the outflow velocity to $\sim 1000 -2000$~\kmps for both AGN or starburst-driven winds \citep[e.g.][]{mur05,thom15}.
Therefore, the incidence of outflows in AGN identified at different wavelengths may reveal information about the potential mechanism(s) driving these outflows.  

We analyze the dependence of the incidence of outflows detected in our sample and the identification wavelength of the AGN.
Outflows are somewhat more commonly detected among X-ray AGN (9/36, $25 \% \pm 8 \%$), 
compared to non-X-ray AGN (4/31, $13 \% \pm 6 \%$).
Similarly, the outflow detection rate is higher among optical AGN (11/45, $24 \% \pm 7 \%$) than 
non-optical AGN (2/22, $9 \% \pm 6 \%$).
There is no difference in the outflow detection rate between IR AGN (5/27, $19 \% \pm 8 \%$) 
and non-IR AGN (8/40, $20 \% \pm 7 \%$).  We note that AGN can be identified at multiple wavelengths;
\citet{aza16} report the overlap and uniqueness of AGN selection at each wavelength in the MOSDEF survey.

The slightly higher outflow incidence in X-ray and/or optical AGN may be due to their bias towards being
detected more often in higher mass galaxies.
It is interesting to note that optically-selected AGN, which reside in more dusty host galaxies in the MOSDEF survey, have a slightly higher outflow incidence than non-optical AGN.
However, no difference is observed between IR AGN and non-IR AGN.
Therefore, we cannot establish a significant relation between the incidence of outflows and 
the dust content of the galaxy or near the AGN, which may be expected if radiation pressure 
on dust grains is a key component of the physical driving mechanism of these winds.
We note that IR AGN in MOSDEF are found to reside in host galaxies which are less dusty on the whole, although there may be more dust in the vicinity of the AGN to allow for IR detection.
Nevertheless, one should be cautious about these conclusions as the current sample size is small.
The full MOSDEF sample will provide better statistics to further investigate these questions.

\subsection{Mass and energy outflow rates}\label{energetics}

\begin{figure*}[!ht]
	\centering
		\includegraphics[width=0.9\textwidth]{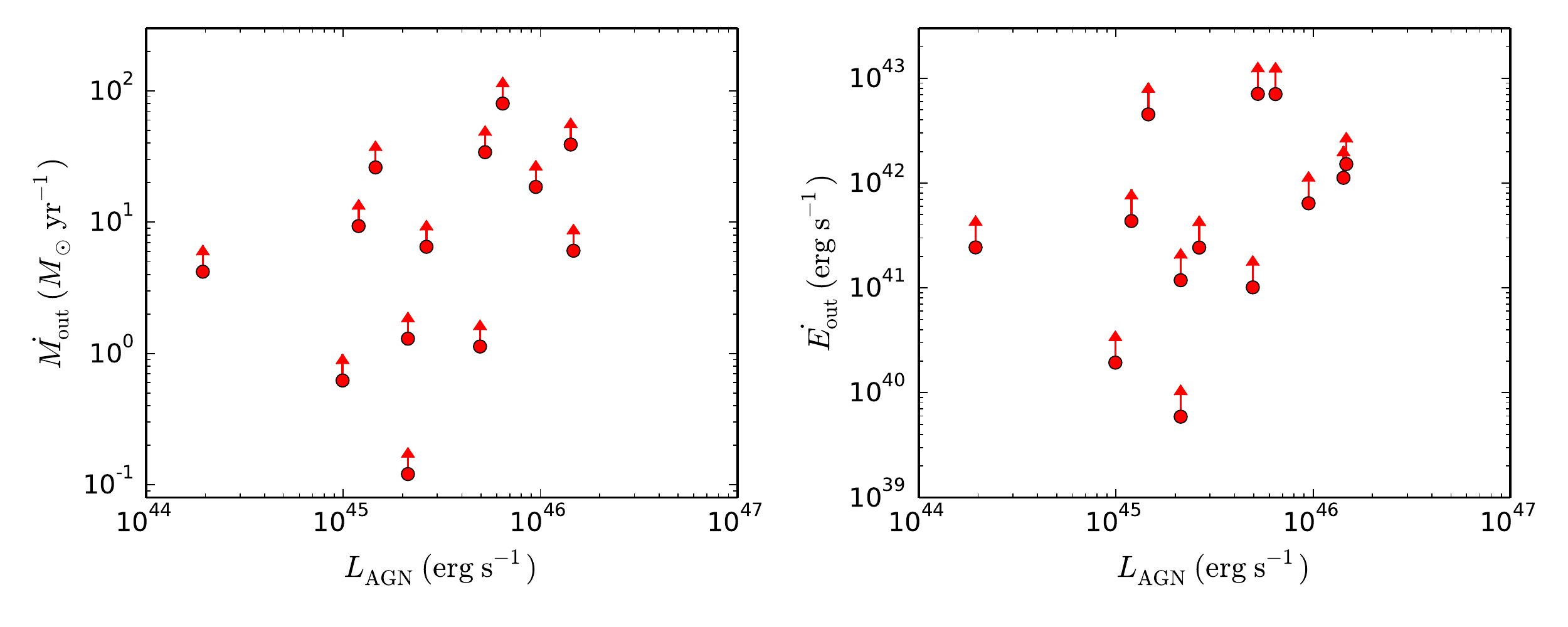}
		\caption{Left: Mass outflow rates estimated for the ionized gas (shown as  lower limits on the total mass outflow rate) versus the bolometric AGN luminosity. Right: Energy outflow rate of the ionized gas (shown as lower limits on the total energy outflow rate) versus the bolometric AGN luminosity.}
	\label{fig:mdot_edot}
\end{figure*}

To understand the drivers of these outflows and their potential impact on the evolution of their 
host galaxies, it is crucial to estimate the mass and energy carried by the outflows.
Unfortunately, deriving accurate estimates of the mass and energy of the outflows can be challenging as it requires precise knowledge of the outflow geometry and kinematics, as well as the properties of the interstellar medium and the physical state of the gas in the outflow.
Nevertheless, first-order estimates of the mass and energy of the outflows are informative to verify the consistency of the proposed picture of AGN-driven outflows quenching star formation.

We estimate the mass of the ionized outflowing gas by counting the number of recombining hydrogen atoms.
Assuming purely photoionized gas with `Case B' recombination with an intrinsic line ratio of  \halpha/\hbeta = 2.9 and an electron temperature of $T = 10^4$K,  following \citet{ost06} and \citet{nes17a}, the mass of the ionized gas in the outflow can be expressed as
\begin{equation}
\frac{M_\mathrm{ion}}{3.3 \times 10^8 \msun} = \left( \frac{L_{\mathrm{H}\alpha,\mathrm{out}}}{10^{43}~\mathrm{erg~s}^{-1}} \right) \left( \frac{n_e}{100~\mathrm{cm}^{-3}} \right)^{-1},
\label{eq:ha}
\end{equation}
where $L_{\mathrm{H}\alpha,\mathrm{out}}$ is the outflow \halpha ~luminosity and $n_e$ is the electron density of the ionized gas.
Equivalently, the mass of the ionized gas in the outflow can be expressed in terms of the \hbeta ~luminosity as
\begin{equation}
\frac{M_\mathrm{ion}}{9.5 \times 10^8 \msun} = \left( \frac{L_{\mathrm{H}\beta,\mathrm{out}}}{10^{43}~\mathrm{erg~s}^{-1}} \right) \left( \frac{n_e}{100~\mathrm{cm}^{-3}} \right)^{-1},
\label{eq:hb}
\end{equation}
where $L_{\mathrm{H}\beta,\mathrm{out}}$ is the outflow \hbeta ~luminosity. As noted in \citet{nes17a}, these updated equations are a factor of 3 smaller than in some previous studies \citep[e.g.][]{nes06,nes11}.
In 11 of the 14 outflows (we separate the two outflows detected in COSMOS 11223, such that we have a total of 14 outflows), both \halpha ~and \hbeta ~are measured.
For all of these outflows, the signal-to-noise in \halpha ~is higher than that in \hbeta , so Equation \ref{eq:ha} is used.
Equation \ref{eq:ha} is also used for GOODS-N 8482, where only \halpha ~is measured.
Only \hbeta ~is available in UDS 6055, and equation \ref{eq:hb} is used.
Neither \halpha ~nor \hbeta ~is measured in AEGIS 31250, so no mass estimate is calculated.
In our sample, the ionized masses for the same source obtained from Equations \ref{eq:ha} and \ref{eq:hb} are in good agreement, differing by at most a factor of five.

The electron density of the {\it outflow} is not directly measured in MOSDEF as the outflow component is not detectable in the relatively faint [SII] doublet.  
However, the median electron density in the {\it star-forming regions} of MOSDEF galaxies at $z \sim 2.3$ is $n_e = 250^{+128}_{-29}~\mathrm{cm}^{-3}$ \citep{san16}.
A recent study \citep{new12} suggests that the outflowing gas density is probably a factor of a few smaller than that of the star-forming gas.
Therefore, we take the electron density of the outflow as $n_e = 100~\mathrm{cm}^{-3}$ when calculating the ionized outflow mass in Equations \ref{eq:ha} and \ref{eq:hb}.
While there is uncertainty in this value, this is similar to the electron densities used in other studies of AGN outflows \citep[e.g.][]{nes06,nes08,liu13b,gen14, har14}, which lie between 80 and 500 $\mathrm{cm}^{-3}$.
Using this method, we obtain ionized outflowing gas masses of $M_\mathrm{ion} = (0.5-34)\times 10^7 \msun$, with a median mass of $4.3\times 10^7 \msun$.

The kinetic energy of the ionized outflow can be estimated by
\begin{equation}
E_\mathrm{kin} = \frac{1}{2} M_\mathrm{out} v_\mathrm{out}^2.
\end{equation}
We use $v_\mathrm{out} = v_\mathrm{max}$ as defined in Section \ref{kin} for the outflow velocity. 
And from here onwards, we use $M_\mathrm{out} = M_\mathrm{ion}$.
For our sources we find $E_\mathrm{kin} = (0.07-5.6) \times 10^{56}$ erg, with a median energy of $1.2 \times 10^{56}$ erg.

Then the mass outflow rate is obtained by
\begin{equation}
\dot{M}_\mathrm{out} = M_\mathrm{out} \times \frac{v_\mathrm{out}}{R_\mathrm{out}}.
\end{equation}
The value of $R_\mathrm{out}$ is taken to be $r_{10}$ as defined in Section \ref{ext}, which is the distance where the flux of the outflow as seen in emission drops to one tenth of the maximum value.
The kinetic energy outflow rate and momentum flux are then given by
\begin{equation}
\dot{E}_\mathrm{kin} = \frac{1}{2} \dot{M}_\mathrm{out} v_\mathrm{out}^2
\end{equation} and
\begin{equation}
\dot{P}_\mathrm{out} = \dot{M}_\mathrm{out} v_\mathrm{out},
\end{equation}
respectively.
We find $\dot{M}_\mathrm{out} = 0.1-80.0~\msun~\mathrm{yr}^{-1}$, with a median mass outflow rate of $6.5~\msun~\mathrm{yr}^{-1}$.
The kinetic energy outflow rate is $\dot{E}_\mathrm{kin} = (0.06-70.9) \times 10^{41} ~\mathrm{erg~s}^{-1}$, with a median of $4.3 \times 10^{41} ~\mathrm{erg~s}^{-1}$.
The momentum flux is $\dot{P}_\mathrm{out} = (0.03-27) \times 10^{34}~\mathrm{dyn}$, with a median of $2.3 \times 10^{34} ~\mathrm{dyn}$.
Values of the mass and energy outflow rates as a function of bolometric AGN luminosity are shown in Figure \ref{fig:mdot_edot} and listed in Table \ref{table3}.

\capstartfalse
\begin{deluxetable*}{lrrrrrrr}[ht!]
\centering
\tablecaption{Ionized Outflow Energetics Measurements}
\tablehead{
\colhead{ID}&\colhead{$\dot{M}$\tablenotemark{a}}&
\colhead{$\dot{M}/\dot{M}_\mathrm{stellar,max}$\tablenotemark{b}}&
\colhead{$\dot{M}/\mathrm{SFR}$\tablenotemark{c}}&
\colhead{$\dot{E}$\tablenotemark{d}}&
\colhead{$\dot{E}/L_\mathrm{AGN}$\tablenotemark{e}}&
\colhead{$\dot{P}$\tablenotemark{f}}&
\colhead{$\dot{P}c/L_\mathrm{AGN}$\tablenotemark{g}}
}
\startdata
6055	&	80.04	&	13.56	&	30.43	&	7.06$\times 10^{42}$	&	0.11\%	&	2.67$\times 10^{35}$	&	1.24	\\
20891	&	39.04	&	0.43	&	0.30	&	1.12$\times 10^{42}$	&	0.01\%	&	7.43$\times 10^{34}$	&	0.16	\\
15286	&	26.13	&	0.04	&	0.01	&	4.52$\times 10^{42}$	&	0.31\%	&	1.22$\times 10^{35}$	&	2.51	\\
20979	&	18.56	&	0.28	&	0.22	&	6.41$\times 10^{41}$	&	0.01\%	&	3.87$\times 10^{34}$	&	0.12	\\
11223.2	&	1.30	&	0.01	&	0.01	&	1.18$\times 10^{41}$	&	0.01\%	&	4.40$\times 10^{33}$	&	0.06	\\
11223.3	&	0.12	&	0.00	&	0.00	&	5.91$\times 10^{39}$	&	0.00\%	&	3.00$\times 10^{32}$	&	0.00	\\
9367	&	6.06	&	0.57	&	0.98	&	1.52$\times 10^{42}$	&	0.01\%	&	3.41$\times 10^{34}$	&	0.07	\\
8482	&	0.62	&	0.01	&	0.01	&	1.94$\times 10^{40}$	&	0.00\%	&	1.23$\times 10^{33}$	&	0.04	\\
6976	&	1.13	&	0.25	&	0.64	&	1.01$\times 10^{41}$	&	0.00\%	&	3.80$\times 10^{33}$	&	0.02	\\
10421	&	34.12	&	2.06	&	2.97	&	7.09$\times 10^{42}$	&	0.14\%	&	1.75$\times 10^{35}$	&	1.00	\\
30014	&	4.20	&	0.54	&	1.08	&	2.44$\times 10^{41}$	&	0.13\%	&	1.14$\times 10^{34}$	&	1.75	\\
17754	&	9.34	&	0.22	&	0.21	&	4.34$\times 10^{41}$	&	0.04\%	&	2.26$\times 10^{34}$	&	0.57	\\
17664	&	6.50	&	0.22	&	0.25	&	2.42$\times 10^{41}$	&	0.01\%	&	1.41$\times 10^{34}$	&	0.16	

\enddata
\tablenotetext{a}{Ionized mass outflow rates, in units of $\msun ~\mathrm{yr}^{-1}$.}
\tablenotetext{b}{Ratios of the estimated ionized mass outflow rate to the maximum theoretically-predicted value from stellar feedback.}
\tablenotetext{c}{Ionized mass-loading factors of the outflows.}
\tablenotetext{d}{Energy outflow rates of the ionized outflow, in units of erg s$^{-1}$.}
\tablenotetext{e}{Ratios of the measured energy outflow rate to the bolometric luminosity of the AGN.}
\tablenotetext{f}{Momentum flux, in units of dyn.}
\tablenotetext{g}{Ratios of the measured momentum flux to the radiation pressure of the AGN.}

\label{table3}
\end{deluxetable*}

These estimates provide lower limits on the actual outflow rates as they consider only the ionized phase of the outflowing gas, ignoring outflowing gas in other phases (e.g. molecular and neutral).
For example, \citet{nes17a} find that that there is $\sim5-50$ times more molecular gas
than ionized gas in high-redshift radio galaxies with outflows.
Moreover, the observed \halpha ~or \hbeta ~flux does not account for the entire mass of the ionized outflows, since we observe only the blueshifted wing in the emission line.
Additionally, our estimates are derived assuming $n_e = 100~\mathrm{cm}^{-3}$.
Since the ionized gas mass is inversely proportional to the value of $n_e$, any change in the value of $n_e$ will inversely change both the mass and energy outflow rates.
For example, if the value of $n_e$ decreases by a factor of ten, the mass and energy outflow rates will increase by a factor of ten.
Therefore, it is reasonable that the actual outflow mass is a factor of at least a few to ten times larger the the ionized gas mass estimated here, and these number can serve as a conservative lower limit.

Another method commonly used in the literature to estimate mass and energy outflow rates is to consider an energy conserving bubble expanding into a uniform medium \citep[e.g.][]{hec90, vei05, nes06, nes08, har12, har14}.
In this scenario, the energy outflow rate is estimated by 
\begin{equation}\label{eq:e_con}
\dot{E}_\mathrm{out} = 1.5 \times 10^{35} \left(\frac{r}{\mathrm{kpc}}\right)^2 \left(\frac{v}{\mathrm{km~s}^{-1}}\right)^3 n_{0.5} ~\mathrm{erg~s}^{-1},
\end{equation}
where $n_{0.5}$ is the ambient density in units of  $0.5~\mathrm{cm}^{-3}$.
If we use the observed values of $r$ and $v$ for our outflows and assume $n_{0.5}$ to 
be unity, we obtain energy injection rates that are $\sim 200$ times higher than those obtained above by counting recombining hydrogen atoms.
As this energy-conserving method results in unreasonably large estimates of the mass and energy outflow rates, we do not employ it further in this paper.

As a sanity check of the outflow masses estimated above by counting recombining hydrogen atoms, we compare the derived cooling luminosity of the outflows ($L_\mathrm{cool}$) to the AGN luminosity.
A realistic outflow mass should produce a cooling luminosity that does not exceed the AGN luminosity, if the gas in the outflows is indeed photoionized by the AGN.
The cooling luminosity is given by 
\begin{equation}
\begin{aligned}
L_\mathrm{cool} &= \int \Lambda(T) n_e^2 dV  \\
&= 1.19 \times 10^{44} ~\mathrm{erg~s^{-1}} n_{e,2} M_\mathrm{out,8} \Lambda_{-23} (T)
\label{eq:lcool},
\end{aligned}
\end{equation}
where $n_{e,2}$ is the electron density in units of $100~\mathrm{cm^{-3}}$,
$M_\mathrm{out,8}$ is the outflow mass in units of $10^8 \msun$ and
$\Lambda_{-23} (T)$ is the cooling function in units of $10^{-23}~\mathrm{erg~cm^3~s^{-1}}$, taken to be 1 at $10^4$~K, the likely temperature of the ionized gas.
We find that the cooling luminosity is much less than the AGN luminosity for all the outflows from our estimates, with an average ratio of $\gtrsim 0.03$.
The ratios obtained using the energy conserving approximation (Eq. \ref{eq:e_con}), however, have a median value of $\sim 60$, indicating that the mass outflow rates estimated using this method are physically unrealistic.

\subsection{Physical driver of outflows and impact on host galaxies}

\begin{figure}[!t]
	\centering
		\includegraphics[width=0.5\textwidth]{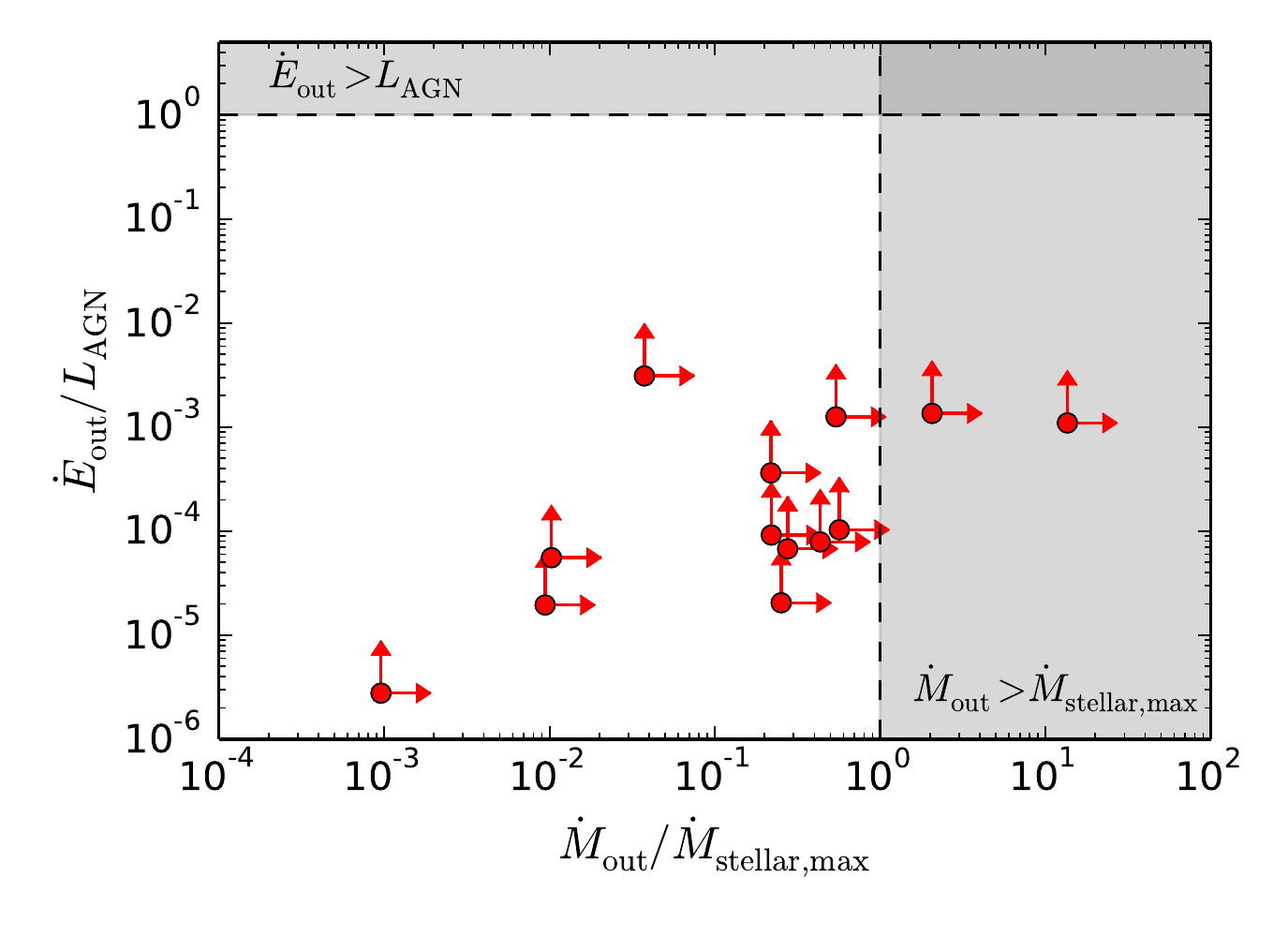}
		\caption{The ratio of the kinetic energy rates in the outflow to the AGN luminosity versus the ratio of the observed mass outflow rates to the maximum mass outflow rates that could be provided by stellar feedback, given the SFR of the host galaxy \citep[see][]{hop12}. The shaded area at the top indicates where the kinetic energy outflow rates exceed the AGN luminosity (such that AGN could not drive the outflows), while the shaded area on the right indicates where the observed mass outflow rates exceed the maximum mass outflow rates that could be due to stellar feedback (such that stars could not drive the outflows).}
	\label{fig:ratio}
\end{figure}

To investigate the possible drivers of the outflows, it is common to compare the mass and energy available from potential drivers with that carried by the outflows.
The bolometric luminosity of the AGN ($L_\mathrm{AGN}$) provides a measurement of the energy available from the AGN to drive the outflows.
The bolometric luminosity of the AGN is estimated by applying a bolometric correction of 600 times the [OIII]$\lambda 5008$ luminosity \citep{kau09}.
We also consider another important driver of galaxy-wide outflows, stellar feedback, and the mass potentially available from it.
\citet{hop12} provides a mass loss rate by stellar feedback as a function of star formation rate given by
\begin{equation}
\dot{M}_\mathrm{stellar, max} = 3 \msun ~\mathrm{yr}^{-1} \left( \frac{\mathrm{SFR}}{\msun ~\mathrm{yr}^{-1}} \right)^{0.7}.
\end{equation}
This mass loss rate improves on earlier work, in that it considers not only direct mass loss from supernovae and stellar winds, but also the subsequent entrainment of the interstellar medium.
Moreover, it includes all material that is being ejected out of the galaxy, in all phases, locations and directions, which tends to overestimate what is actually observed.
Therefore, it can be considered as an upper limit to the observed mass outflow rates.
This maximum mass loading factors of stellar feedback ($\dot{M}_\mathrm{stellar, max}/\mathrm{SFR}$) for our AGN host galaxies are of the order of unity.

In Figure \ref{fig:ratio} we compare the kinetic energy rates of the ionized outflows with $L_\mathrm{AGN}$ and the ionized mass outflow rates to the maximum possible from  stellar feedback.
The energy outflow rates range from $>0.0003-0.3 \%$ of $L_\mathrm{AGN}$, with a median of $>0.009 \%$.  This shows that the AGN are energetically capable of driving the outflows.  
The ionized mass outflow rates range from $>0.1-1400 \%$ of the maximum mass loss rate from stellar feedback, with a median of $>25 \%$.
Two of the outflows have mass outflow rates lower limits exceeding the maximum mass loss rates of stellar feedback, indicating that stellar feedback must not be the sole driver of these two outflows, and an alternative driver is required.
It is important to note that the value of $\dot{M}_\mathrm{stellar, max}$ represents an upper limit to the observed mass outflow rate from stellar feedback, while the value of $\dot{M}_\mathrm{out}$ is a strict lower limit to the total mass outflow rate.
The actual values of $\dot{M}_\mathrm{out}/\dot{M}_\mathrm{stellar, max}$ are likely  larger by an order of magnitude; this is a conservative indicator of the need of an alternative driver.
Therefore, stellar feedback is likely not the sole driver of these outflows, and an alternative driver, such as AGN, is required.

\begin{figure*}[!ht]
	\centering
		\includegraphics[width=0.9\textwidth]{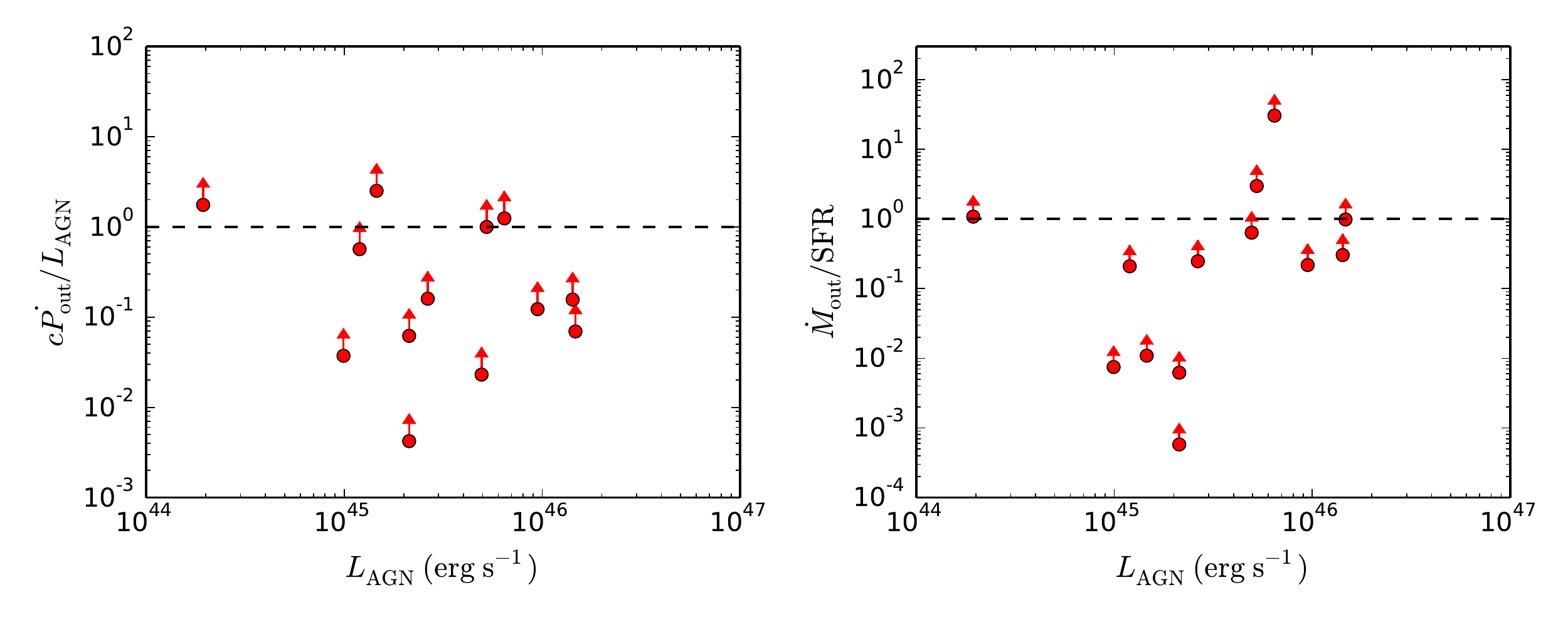}
		\caption{The ratio of the momentum flux of the outflow to the radiation pressure of the AGN versus the luminosity of the AGN. 
The outflow momentum flux exceeds the radiation pressure of the AGN for a considerable number of sources, suggesting radiation pressure alone is insufficient to drive the outflows.
The mass loading factors of the outflows versus the bolometric AGN luminosity.
The median mass loading factor lower limit is $>0.24$.}
	\label{fig:pdot_eta}
\end{figure*}

One possible mechanism for the AGN to drive the outflow is by its radiation pressure on dust grains in the surrounding material \citep[e.g.][]{mur05,thom15}.
In this case, the outflow is momentum-driven, and it is useful to compare the momentum flux of the outflow to the radiation pressure of the AGN to test this scenario.
 In the left panel of Figure \ref{fig:pdot_eta}, we plot the ratio of the momentum flux of the outflow ($\dot{P} = \dot{M} v$) to the radiation pressure of the AGN ($L_\mathrm{AGN}/c$) as a function of the AGN bolometric luminosity. 
Three of the outflows have lower limits of the momentum ratio greater than unity, with a median momentum ratio of $>0.16$.
This shows that most of these outflows can be driven by radiation pressure alone.  
However, given that the measured momentum fluxes are strict lower limits, as we are observing the ionized phase only, if the total values are an order of magnitude higher  then another mechanism must be involved.  
Additionally, the momentum flux of the outflows could be boosted by a factor of $\sim 15$ by work done by the hot post-shock gas during the energy-conserving phase \citep{fau12}.
This suggests that for at least some of these sources, radiation pressure on dust
grains may be insufficient to drive these outflows.

An important measure of the power and impact of the outflow on the host galaxy is the mass loading factor, $\eta = \dot{M}_\mathrm{out}/\mathrm{SFR}$, which compares
the mass outflow rate of gas being ejected from the galaxy to the rate at which gas is 
being converted into stars within the galaxy.  For our soures, lower limits on $\eta$ range from $>0.01$ to $>30$, with a median of $>0.24$.  The large mass loading factor for UDS 6055 is due to the low SFR measured for this galaxy.
The mass loading factors of the outflows are shown in the right panel of Figure \ref{fig:pdot_eta}.
Given that these estimates are strict lower limits, the actual mass loading factor is likely a factor of a few to ten times higher, equal to or higher than unity.
Such high mass loading factors imply that these outflows are at least capable of regulating star formation in the host galaxies, although it is unclear whether they can remove sufficient gas to quench star formation.
This figure shows that there may by an AGN luminosity threshold of $10^{45.5} ~\mathrm{erg~s}^{-1}$, above which the mass loading factor is greater than unity, 
though a larger sample is required to determine whether this is significant.

\subsection{Positive AGN Feedback in MOSDEF and SDSS}\label{pos}

\begin{figure*}[!ht]
	\centering
		\includegraphics[width=\textwidth]{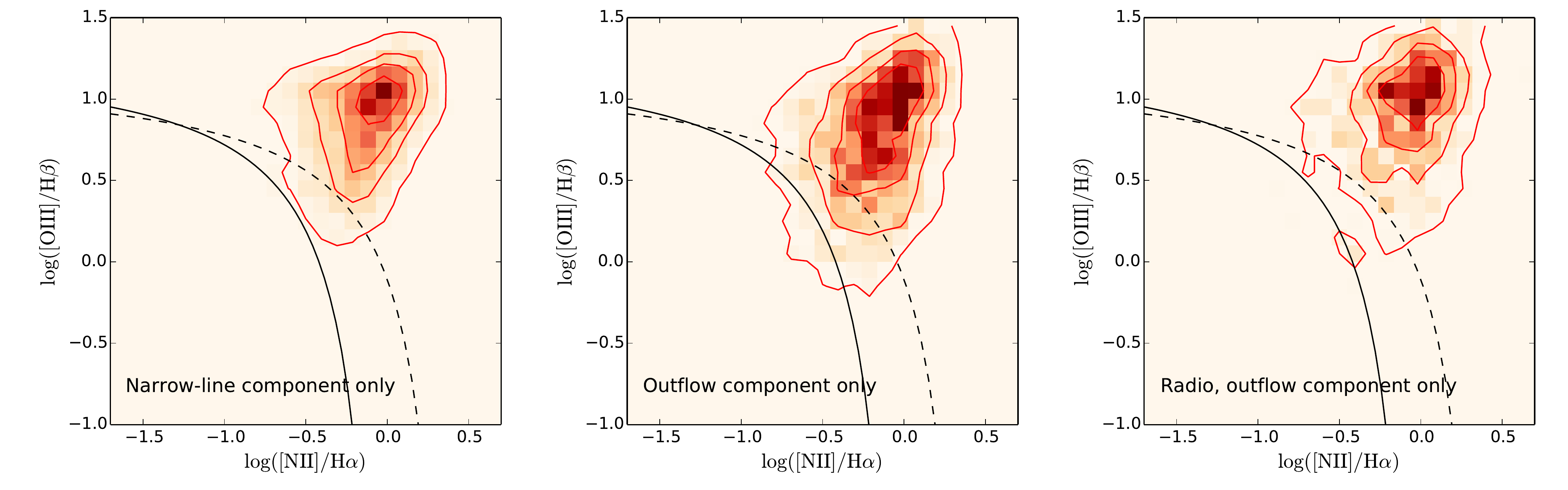}
		\caption{BPT diagrams for optically-selected AGN with outflows in SDSS using line ratios from  \citet{mul13}. Only sources with signal-to-noise greater than 3 in all four of the relevant lines used for this diagram are shown here. The left panel shows the line ratios for the narrow-line component only, containing 4881 sources. The middle panel shows the line ratios for gas in the outflow component only, containing 2705 sources. The right panel shows the line ratios for the outflow component only for sources from the middle panel which are also detected in the radio, using the FIRST survey, containing 759 sources.
There is no increase in the fraction of sources below the \citet{kau03} line for the outflow component in radio-loud sources compared with the middle panel, as may be expected if positive AGN feedback is a widespread phenomenon.
}
	\label{fig:positive}
\end{figure*}

AGN feedback is typically invoked as a possible mechanism to quench star formation in galaxies.  This is referred to as ``negative'' feedback, in that star formation is reduced or shut off entirely.  Another possibility has been raised, which is that AGN feedback may instead {\it trigger} star formation; this is referred to as ``positive'' feedback, as it leads to an enhancement in star formation.  

The motivation for positive AGN feedback originated in observations of a handful of elliptical galaxies at low redshift that have clear radio jets from AGN, for which regions of star formation are clearly seen along the jet \citep[e.g.,][]{Croft06, Feain07,Inskip08}.  Additionally, at high redshift it has been observed that the axis of radio AGN jets was often aligned with an elongation in the surface density of star formation \citep[e.g.,][ and references therein]{Bicknell00}.  Recently, several theoretical papers have studied this idea using hydrodynamical simulations of galaxies, in which an AGN-driven jet is seen to be able to induce star formation in a ring in the host galaxy, which propagates outward \citep[e.g.,][]{Silk05,Gaibler12, Ishibashi12}.  It has been speculated that such triggered star formation could substantially affect the growth of galaxies \citep{Ishibashi13}. 

A clear observational signature of such positive AGN feedback would be investigating radio AGN with jets and using the BPT diagram to detect star formation in the outflowing gas, along the jet.  Indeed, for the well-studied source known as Minkowski's Object, the gas that appears to be compressed by the jet does have line ratios in the BPT diagram in the location of the star-forming sequence, well below the AGN region of the diagram \citep{Croft06}.  This argues that the jet has induced star formation in this gas. 

In the MOSDEF sample, we present the locations of outflowing gas from eight sources in the right panel of Figure \ref{fig:BPT} and find that the locations lie in the AGN region; this outflowing gas does not appear to be photoionized by star formation.  This would argue against positive AGN feedback being a widespread phenomenon at $z\sim2$.  However, our sample size is very small, and only two of our sources are known to be radio-loud, which in the observations at both low and high redshift appears to be a prerequisite for positive AGN feedback.  Clearly, a much larger sample of radio AGN would be required to robustly test this scenario.

The SDSS provides such a sample at low redshift.  To investigate this, we use the publicly-available optical AGN sample of \citet{mul13}, who use the BPT diagram to identify $\sim$25,000 AGN in SDSS, using the line ratios from the SDSS spectral pipeline.  This pipeline does not include multiple kinematic components for each line, which \citet{mul13} subsequently perform.  They identify 9,637 sources with a significant (S/N$>$3) blueshifted kinematic component in the [OIII] 5007 \AA ~line, in addition to the narrow line observed at the systemic redshift of the galaxy.

In Figure \ref{fig:positive}, we show in the left panel the {\it narrow-line} only line ratios for these optical AGN with outflows, while the middle panel shows the line ratios for the {\it outflow} component only.  The narrow-line ratios almost exclusively lie above the \citet{kau03} line (only 3\% of sources are below this), while the outflow component line ratios lie somewhat lower, with 6\% of the outflow line ratios below the \citet{kau03} line.  The difference between these panels may argue that positive AGN feedback occurs in a few percent of these sources, however, the more important test is whether the line ratios shift downward for AGN with jets, i.e., radio AGN.  In the right panel, therefore, we show the line ratios for the AGN in the \citet{mul13} sample that have outflows detected in the [OIII] line and are also detected in the FIRST survey \citep{bec95}.  Here we find that only 3\% of these sources lie below the \citet{kau03} line, consistent with the left panel.  This argues that positive AGN feedback is not observed to be a widespread phenomenon at $z\sim0.1$.  For it to be a common phenomenon, it must then occur only over a relatively short timescale, or it may just be a rare phenomenon.

\section{Conclusions}\label{conc}

Using data from the first two years of the MOSDEF survey, we analyze the rest frame optical spectra of 67 X-ray, IR and/or optically-selected AGN at $z \sim 2$ ~to identify and characterize AGN-driven outflows at $z \sim 2$.
We identify outflows from the presence of an additional blueshifted component in the profiles of the \hbeta , [OIII], \halpha ~and [NII] emission lines, and we remove sources which appear to be ongoing mergers from their {\it HST} morphologies.
The bolometric luminosities of the AGN in this study span $10^{44}-10^{46}~\mathrm{erg~s^{-1}}$, including both quasars and more common, moderate-luminosity AGN.
Our main conclusions are as follows.

\begin{enumerate}
\item 
The velocities of the outflows range from 300 to 1000 km s$^{-1}$. 
Eight out of the 13 detected outflows are spatially extended, having spatial extents of 2.5 to 11.0 kpc along the MOSFIRE slits. 
\item 
Outflows are detected in 13 (19\%) out of 67 AGN. 
This can be considered a lower limit on the true incidence as relatively high signal-to-noise spectra are required to detect the outflows.
By contrast, using the same analysis procedures, outflows are detected in only eight (1.8\%) out of the 457 MOSDEF galaxies.
The outflow detection rate increases with the [OIII] luminosity of the AGN, but the same trend is also observed as the signal-to-noise ratio of the spectra increases, as there is a strong correlation between $L_\mathrm{[OIII]}$ and the signal-to-noise of the spectra.
\item 
Outflows are detected across the galaxy star forming main sequence spanning stellar masses of $10^{10}-10^{11.5} \msun$.
No significant trend of incidence with stellar mass is observed.
The incidence of outflows also does not depend on the SFR of the host galaxy with respect to the main sequence.
\item 
Using line ratio diagnostics in the BPT diagram for the narrow-line and outflow components separately, we find that the line ratios of the outflowing gas are shifted towards the AGN region of the BPT diagram.  
This indicates the the outflowing gas is photoionized by the AGN.
\item 
The line ratio diagnostics in the outflows argues against ``positive'' AGN feedback, in which the presence of the outflow triggers, as opposed to suppresses, star formation.  
As the MOSDEF AGN outflow sample is small, we also analyzed line ratios for AGN outflows in SDSS, specifically investigating radio-loud AGN, and again found little evidence for positive AGN feedback being a widespread or long-lasting phenomenon.
\item 
The typical ionized mass outflow rate of our sources is $\gtrsim 10 \msun ~\mathrm{yr^{-1}}$, while the typical energy rate of the ionzied gas is $\gtrsim 10^{42}~\mathrm{erg~s^{-1}}$.
These lower limits on the total mass outflow rates are typically at least comparable to the theoretically-predicted maximum values from stellar feedback, while the energy outflow rates are $\sim 0.01\%$ the bolometric AGN luminosities. 
This indicates the stellar feedback is likely insufficient to drive all of these outflows, 
while the energetic output of the AGN is more than sufficient to drive them. 
Taken together with our finding that a galaxy is 10 times more likely to have a detected outflow if it hosts an AGN, it appears very likely that these outflows are AGN-driven. 
\item 
The mass loading factors of these outflows are on the order of unity, which suggests that the outflows likely help to regulate star formation in the host galaxies but may not be sufficient to fully quench star formation.
\end{enumerate}

Our results show that galaxy-wide AGN-driven outflows are common not only for quasars but also for moderate-luminosity AGN at $z \sim 2$.
Theoretical studies of galaxy formation and evolution should therefore account for AGN feedback from moderate-luminosity AGN, in addition to feedback from quasars.
In the future, the full MOSDEF data will provide a larger sample to better study correlations between AGN outflow properties and host galaxy or AGN properties at $z \sim 2$.  
This study and future similar AGN outflow studies from large galaxy samples also provide targets for more detailed IFU investigations into the physical nature of AGN-driven outflows at the peak of cosmic galaxy growth.

We thank Norm Murray,  Patrick Diamond and Claude-Andr{\`e} Faucher-Gigu{\`e}re for useful discussions.
We also thank the anonymous referee for helping to improve the paper.
This work would not have been possible without the 3D-{\it HST} collaboration, who provided us the spectroscopic and photometric catalogs used to select our targets and to derive stellar population parameters.
This work is partially supported by NSF CAREER grant 1055081, awarded to A. Coil.
Funding for the MOSDEF survey is provided by NSF AAG grants AST-1312780, 1312547, 1312764, and 1313171, and archival grant AR-13907, provided by NASA through a grant from the Space Telescope Science Institute.
The data presented herein were obtained at the W.M. Keck Observatory, which is operated as a scientific partnership among the California Institute of Technology, the University of California and the National Aeronautics and Space Administration. 
The Observatory was made possible by the generous financial support of the W.M. Keck Foundation. 
The authors wish to recognize and acknowledge the very significant cultural role and reverence that the summit of Mauna Kea has always had within the indigenous Hawaiian community. 
We are most fortunate to have the opportunity to conduct observations from this mountain.

\bibliographystyle{apj}

\begin{thebibliography}{}
\expandafter\ifx\csname natexlab\endcsname\relax\def\natexlab#1{#1}\fi

\bibitem[Aird et al.(2012)]{aird12}
	Aird, J., Coil, A.~L., Moustakas, J., et al.\ 2012, \apj, 746, 90
\bibitem[Aird et al.(2015)]{aird15}
	Aird, J., Coil, A.~L., Georgakakis, A., et al.\ 2015, \mnras, 451, 1892 
\bibitem[Antonucci(1993)]{ant93}
	Antonucci, R. 1993, \araa, 31, 473
\bibitem[Antonuccio-Delogu \& et al.(2010)]{ant10} 
	Antonuccio-Delogu, V., \& Silk, J.\ 2010, AGN Feedback in Galaxy Formation
\bibitem[Azadi et al.(2017)]{aza16} 
	Azadi, M., Coil, A.~L., Aird, J., et al.\ 2017, \apj, 835, 27 
\bibitem[Baldwin et al.(1981)]{bal81}
	Baldwin, J.~A., Phillips, M.~M., \& Terlevich, R.\ 1981, \pasp, 93, 5
\bibitem[Becker et al.(1995)]{bec95} 
	Becker, R.~H., White, R.~L., \& Helfand, D.~J.\ 1995, \apj, 450, 559 
\bibitem[Benson et al.(2003)]{ben03}
	Benson, A. J., Bower, R. G., Frenk, C. S., et al. 2003, \apj, 599, 38
\bibitem[Bicknell et al.(2000)]{Bicknell00} 
	Bicknell, G.~V., Sutherland, R.~S., van Breugel, W.~J.~M., et al.\ 2000, \apj, 540, 678 
\bibitem[Brammer et al.(2012)]{bram12}
	Brammer, G.~B., van Dokkum, P.~G., Franx, M., et al.\ 2012, \apjs, 200, 13 
\bibitem[Brusa et al.(2007)]{bru07}
	Brusa, M., Zamorani, G., Comastri, A., et al.\ 2007, \apjs, 172, 353
\bibitem[Brusa (2015)]{bru15}
	Brusa, M., Bongiorno, A., Cresci, G., et al. 2015, \mnras, 446, 2394
\bibitem[Calzetti et al.(2000)]{cal00}
	Calzetti, D., Armus, L., Bohlin, R.~C., et al.\ 2000, \apj, 533, 682
\bibitem[Chabrier(2003)]{cha03}
	Chabrier, G.\ 2003, \pasp, 115, 763
\bibitem[Ciliegi et al.(2003)]{cil03}
	Ciliegi, P., Zamorani, G., Hasinger, G., et al.\ 2003, \aap, 398, 901
\bibitem[Ciliegi et al.(2005)]{cil05}
	Ciliegi, P., Zamorani, G., Bondi, M., et al.\ 2005, \aap, 441, 879
\bibitem[Coil et al.(2015)]{coil15}
	Coil, A.~L., Aird, J., Reddy, N., et al.\ 2015, \apj, 801, 35
\bibitem[Conroy et al.(2009)]{con09}
	Conroy, C., Gunn, J.~E., \& White, M.\ 2009, \apj, 699, 486
\bibitem[Croft et al.(2006)]{Croft06} 
	Croft, S., van Breugel, W., de Vries, W., et al.\ 2006, \apj, 647, 1040 
\bibitem[Debuhr et al.(2012)]{deb12}
	Debuhr, J., Quataert, E., \& Ma, C.-P. 2012, \mnras, 420, 2221
\bibitem[Di Matteo et al.(2005)]{dm05}
	Di Matteo, T., Springel, V., \& Hernquist, L.\ 2005, \nat, 433, 604
\bibitem[Donley et al.(2012)]{don12}
	Donley, J.~L., Koekemoer, A.~M., Brusa, M., et al.\ 2012, \apj, 748, 142
\bibitem[Faucher-Gigu{\`e}re \& Quataert(2012)]{fau12}
	Faucher-Gigu{\`e}re, C.-A., \& Quataert, E.\ 2012, \mnras, 425, 605 
\bibitem[Feain et al.(2007)]{Feain07} 
	Feain, I.~J., Papadopoulos, P.~P., Ekers, R.~D., \& Middelberg, E.\ 2007, \apj, 662, 872 
\bibitem[Ferrarese \& Merritt(2000)]{fer00}
	Ferrarese, L., \& Merritt, D. 2000, \apj, 539, L9
\bibitem[Freeman et al.(in prep)]{free17}
	Freeman, W.~R., Siana, B., Kriek, M., et al.\ in prep
\bibitem[Gaibler et al.(2012)]{Gaibler12} 
	Gaibler, V., Khochfar, S., Krause, M., \& Silk, J.\ 2012, \mnras, 425, 438 
\bibitem[Gabor \& Bournaud(2014)]{gar14} 
	Gabor, J.~M., \& Bournaud, F.\ 2014, \mnras, 441, 1615 
\bibitem[Gebhardt et al.(2000)]{geb00}
	Gebhardt, K., Bender, R., Bower, G., et al. 2000, \apj, 539, L13
\bibitem[Genzel et al.(2014)]{gen14}
	Genzel, R., F{\"o}rster Schreiber, N.~M., Rosario, D., et al.\ 2014, \apj, 796, 7
\bibitem[Georgakakis et al.(2014)]{geo14}
	Georgakakis, A., P{\'e}rez-Gonz{\'a}lez, P.~G., Fanidakis, N., et al.\ 2014, \mnras, 440, 339
\bibitem[Gnedin \& Hollon(2012)]{gne12} 
	Gnedin, N.~Y., \& Hollon, N.\ 2012, \apjs, 202, 13 
\bibitem[Grogin et al.(2011)]{gro11}
	Grogin, N.~A., Kocevski, D.~D., Faber, S.~M., et al.\ 2011, \apjs, 197, 35
\bibitem[Harrison et al.(2014)]{har14}
	Harrison, C. M., Alexander, D. M., Mullaney, J. R., \& Swinbank, A. M. 2014, \mnras, 441,3306
\bibitem[Harrison et al.(2016)]{har16}
	Harrison, C.~M., Alexander, D.~M., Mullaney, J.~R., et al.\ 2016, \mnras, 456, 1195
\bibitem[Harrison et al.(2012)]{har12}
	Harrison, C.~M., Alexander, D.~M., Swinbank, A.~M., et al.\ 2012, \mnras, 426, 1073 
\bibitem[Heckman et al.(1990)]{hec90}
	Heckman, T.~M., Armus, L., \& Miley, G.~K.\ 1990, \apjs, 74, 833 
\bibitem[Heckman \& Best(2014)]{hec14}
	Heckman, T. M., \& Best, P. N. 2014, \araa, 52, 589
\bibitem[Hickox et al.(2014)]{hic14} 
	Hickox, R.~C., Mullaney, J.~R., Alexander, D.~M., et al.\ 2014, \apj, 782, 9 
\bibitem[Hopkins \& Beacom(2006)]{hop06b}
	Hopkins, A.~M., \& Beacom, J.~F.\ 2006, \apj, 651, 142
\bibitem[Hopkins et al.(2006)]{hop06a}
	Hopkins, P.~F., Hernquist, L., Cox, T.~J., et al.\ 2006, \apjs, 163, 1
\bibitem[Hopkins et al.(2008)]{hop08}
	Hopkins, P.~F., Hernquist, L., Cox, T.~J., \& Kere{\v s}, D.\ 2008, \apjs, 175, 356-389
\bibitem[Hopkins et al.(2012)]{hop12} 
	Hopkins, P.~F., Quataert, E., \& Murray, N.\ 2012, \mnras, 421, 3522 
\bibitem[Inskip et al.(2008)]{Inskip08} 
	Inskip, K.~J., Villar-Mart{\'{\i}}n, M., Tadhunter, C.~N., et al.\ 2008, \mnras, 386, 1797 
\bibitem[Ishibashi \& Fabian(2012)]{Ishibashi12} 
	Ishibashi, W., \& Fabian, A.~C.\ 2012, \mnras, 427, 2998 
\bibitem[Ishibashi et al.(2013)]{Ishibashi13} 
	Ishibashi, W., Fabian, A.~C., \& Canning, R.~E.~A.\ 2013, \mnras, 431, 2350 
\bibitem[Ivezi{\'c} et al.(2002)]{ive02}
	Ivezi{\'c}, {\v Z}., Menou, K., Knapp, G.~R., et al.\ 2002, \aj, 124, 2364 
\bibitem[Kauffmann et al.(2003)]{kau03}
	Kauffmann, G., Heckman, T.~M., Tremonti, C., et al.\ 2003, \mnras, 346, 1055
\bibitem[Kauffmann \& Heckman(2009)]{kau09} 
	Kauffmann, G., \& Heckman, T.~M.\ 2009, \mnras, 397, 135 
\bibitem[Kennicutt(1998)]{ken98} 
	Kennicutt, R.~C., Jr.\ 1998, \araa, 36, 189 
\bibitem[Kewley et al.(2013a)]{kew13a} 
	Kewley, L.~J., Dopita, M.~A., Leitherer, C., et al.\ 2013a, \apj, 774, 100 
\bibitem[Kewley et al.(2013b)]{kew13}
	Kewley, L.~J., Maier, C., Yabe, K., et al.\ 2013b, \apjl, 774, L10 
\bibitem[Koekemoer et al.(2011)]{koe11}
	Koekemoer, A.~M., Faber, S.~M., Ferguson, H.~C., et al.\ 2011, \apjs, 197, 36
\bibitem[Koratkar \& Blaes(1999)]{kor99}
	Koratkar, A., \& Blaes, O.\ 1999, \pasp, 111, 1 
\bibitem[Kriek et al.(2009)]{kri09}
	Kriek, M., van Dokkum, P.~G., Labb{\'e}, I., et al.\ 2009, \apj, 700, 221 
\bibitem[Kriek et al.(2015)]{kri15}
	Kriek, M., Shapley, A.~E., Reddy, N.~A., et al.\ 2015, \apjs, 218, 15
\bibitem[Lacy et al.(2004)]{lac04}
	Lacy, M., Storrie-Lombardi, L.~J., Sajina, A., et al.\ 2004, \apjs, 154, 166
\bibitem[Laird et al.(2009)]{lai09}
	Laird, E.~S., Nandra, K., Georgakakis, A., et al.\ 2009, \apjs, 180, 102
\bibitem[Liu et al.(2013a)]{liu13}
	Liu, G., Zakamska, N. L., Greene, J. E., et al. 2013, \mnras, 430, 2327
\bibitem[Liu et al.(2013b)]{liu13b}
	Liu, G., Zakamska, N.~L., Greene, J.~E., Nesvadba, N.~P.~H., \& Liu, X.\ 2013, \mnras, 436, 2576 
\bibitem[Luo et al.(2010)]{luo10}
	Luo, B., Brandt, W.~N., Xue, Y.~Q., et al.\ 2010, \apjs, 187, 560
\bibitem[Magorrian et al.(1998)]{mag98}
	Magorrian, J., Tremaine, S., Richstone, D., et al.\ 1998, \aj, 115, 2285
\bibitem[Markwardt(2009)]{mar09} 
	Markwardt, C.~B.\ 2009, Astronomical Data Analysis Software and Systems XVIII, 411, 251
\bibitem[Mel{\'e}ndez et al.(2014)]{mel14} 
	Mel{\'e}ndez, M., Heckman, T.~M., Mart{\'{\i}}nez-Paredes, M., Kraemer, S.~B., \& Mendoza, C.\ 2014, \mnras, 443, 1358 
\bibitem[McLean et al.(2010)]{mcl10}
	McLean, I.~S., Steidel, C.~C., Epps, H., et al.\ 2010, \procspie, 7735, 77351E-77351E-12
\bibitem[McLean et al.(2012)]{mcl12}
	McLean, I.~S., Steidel, C.~C., Epps, H.~W., et al.\ 2012, \procspie, 8446, 84460J
\bibitem[Momcheva et al.(2015)]{mom15}
	Momcheva, I.~G., Brammer, G.~B., van Dokkum, P.~G., et al.\ 2015, arXiv:1510.02106 
\bibitem[Mullaney et al.(2013)]{mul13}
	Mullaney, J. R., Alexander, D. M., Fine, S., et al. 2013, \mnras, 433, 622
\bibitem[Murray et al.(1995)]{mur95} 
	Murray, N., Chiang, J., Grossman, S.~A., \& Voit, G.~M.\ 1995, \apj, 451, 498 
\bibitem[Murray et al.(2005)]{mur05} 
	Murray, N., Quataert, E., \& Thompson, T.~A.\ 2005, \apj, 618, 569 
\bibitem[Nandra et al.(2015)]{nan15}
	Nandra, K., Laird, E.~S., Aird, J.~A., et al.\ 2015, \apjs, 220, 10
\bibitem[Nesvadba et al.(2006)]{nes06}
	Nesvadba, N.~P.~H., Lehnert, M.~D., Eisenhauer, F., et al.\ 2006, \apj, 650, 693 
\bibitem[Nesvadba et al.(2008)]{nes08} 
	Nesvadba, N.~P.~H., Lehnert, M.~D., De Breuck, C., Gilbert, A.~M., \& van Breugel, W.\ 2008, \aap, 491, 407 
\bibitem[Nesvadba et al.(2011)]{nes11}
	Nesvadba, N.~P.~H., Polletta, M., Lehnert, M.~D., et al.\ 2011, \mnras, 415, 2359 
\bibitem[Nesvadba et al.(2017)]{nes17a} 
	Nesvadba, N.~P.~H., De Breuck, C., Lehnert, M.~D., Best, P.~N., \& Collet, C.\ 2017, \aap, 599, A123 
\bibitem[Netzer(2015)]{net15}
	Netzer, H. 2015, \araa, 53, 365
\bibitem[Newman et al.(2012)]{new12} 
	Newman, S.~F., Genzel, R., F{\"o}rster-Schreiber, N.~M., et al.\ 2012, \apj, 761, 43
\bibitem[Newman et al.(2014)]{new14} 
	Newman, S.~F., Buschkamp, P., Genzel, R., et al.\ 2014, \apj, 781, 21 
\bibitem[Osterbrock \& Ferland(2006)]{ost06}
	Osterbrock, D.~E., \& Ferland, G.~J.\ 2006, Astrophysics of gaseous nebulae and active galactic nuclei, 2nd.~ed.~by D.E.~Osterbrock and G.J.~Ferland.~Sausalito, CA: University Science Books
\bibitem[Perna et al.(2015)]{per15}
	Perna, M., Brusa, M., Cresci, G., et al.\ 2015, \aap, 574, A82
\bibitem[Reddy et al.(2015)]{red15} 
	Reddy, N.~A., Kriek, M., Shapley, A.~E., et al.\ 2015, \apj, 806, 259 
\bibitem[Rupke et al.(2005)]{rup05}
	Rupke, D.~S., Veilleux, S., \& Sanders, D.~B.\ 2005, \apjs, 160, 115
\bibitem[Sanders et al.(2016)]{san16}
	Sanders, R.~L., Shapley, A.~E., Kriek, M., et al.\ 2016, \apj, 816, 23
\bibitem[Shapley et al.(2015)]{shap15} 
	Shapley, A.~E., Reddy, N.~A., Kriek, M., et al.\ 2015, \apj, 801, 88 
\bibitem[Shivaei et al.(2015)]{shiv15}
	Shivaei, I., Reddy, N.~A., Shapley, A.~E., et al.\ 2015, \apj, 815, 98 
\bibitem[Silk(2005)]{Silk05} 
	Silk, J.\ 2005, \mnras, 364, 1337 
\bibitem[Skelton et al.(2014)]{skel14}
	Skelton, R.~E., Whitaker, K.~E., Momcheva, I.~G., et al.\ 2014, \apjs, 214, 24
\bibitem[Steidel et al.(2010)]{stei10} 
	Steidel, C.~C., Erb, D.~K., Shapley, A.~E., et al.\ 2010, \apj, 717, 289 
\bibitem[Stern et al.(2005)]{ster05}
	Stern, D., Eisenhardt, P., Gorjian, V., et al.\ 2005, \apj, 631, 163
\bibitem[Thompson et al.(2015)]{thom15}
	Thompson, T.~A., Fabian, A.~C., Quataert, E., \& Murray, N.\ 2015, \mnras, 449, 147 
\bibitem[Ueda et al.(2003)]{ued03}
	Ueda, Y., Akiyama, M., Ohta, K., \& Miyaji, T. 2003, \apj, 598, 886
\bibitem[Veilleux \& Osterbrock(1987)]{vei87}
	Veilleux, S., \& Osterbrock, D.~E.\ 1987, \apjs, 63, 295
\bibitem[Veilleux et al.(2005)]{vei05}
	Veilleux, S., Cecil, G., \& Bland-Hawthorn, J.\ 2005, \araa, 43, 769 
\bibitem[Weedman(1970)]{wee70}
	Weedman, D.~W.\ 1970, \apj, 159, 405 
\bibitem[Woo et al.(2016)]{woo16}
	Woo, J.-H., Bae, H.-J., Son, D., \& Karouzos, M.\ 2016, \apj, 817, 108 
\bibitem[Zakamska \& Greene(2014)]{zak14}
	Zakamska, N.~L., \& Greene, J.~E.\ 2014, \mnras, 442, 784

\end{thebibliography}

\end{document}